\documentclass[showpacs,preprint]{revtex4}

\usepackage{amsmath}
\usepackage{amssymb}
\usepackage{graphicx}
\usepackage{psfrag}

\newcommand{\comm}[2]{\left[#1,#2\right]}

\newcommand{\vac}{\left|\,0\,\right\rangle}
\newcommand{\ket}[1]{\left|#1\right\rangle}
\newcommand{\bra}[1]{\left\langle#1\right|}

\newcommand{\bs}[1]{\boldsymbol{#1}}

\newtheorem{theo}{Theorem}
\newtheorem{lemma}[theo]{Lemma}

\allowdisplaybreaks[4]

\def\ie{\emph{i.e.,\ }}
\def\eg{\emph{e.g.\ }}

\def\b{{\text{b}}}
\def\r{{\text{r}}}
\def\g{{\text{g}}}

\def\y{{\text{y}}}

\begin{document}

\title{Coloron excitations of the SU(3) Haldane--Shastry model}

\author{Dirk Schuricht and Martin Greiter}

\affiliation{Institut f\"ur Theorie der Kondensierten Materie,
  Universit\"at Karlsruhe, Postfach 6980, 76128 Karlsruhe, Germany}

\pagestyle{plain}

\begin{abstract}
  In a recent publication, we proposed two possible wave functions for the
  elementary excitations of the SU(3) Haldane--Shastry model (HSM), but argued
  on very general grounds that only one or the other can be a valid
  excitation.  Here we provide the explicit details of our calculation proving
  that the wave function describing a coloron excitation which transforms
  according to representation $\bar{3}$ under SU(3) rotations if the spins of
  the original model transform according to representation 3, is exact.  We
  further provide an explicit construction of the exact color-polarized
  two-coloron eigenstates, and thereby show that colorons are free but that
  their relative momentum spacings are shifted according to fractional
  statistics with parameter $g=2/3$.  We evaluate the SU(3) spin currents.
  Finally, we interpret our results within the framework of the asymptotic
  Bethe Ansatz and generalize some of them to the case of SU($n$).

\end{abstract}



\maketitle

\section{Introduction}

In recent years, there has been substantial interest in models in
condensed matter physics with symmetry groups larger than SU(2), the
group underlying the usual spin algebra.  In particular,
transition-metal
oxides~\cite{IsobeUeda96,FujiiNakaoYosihamaNishiNakajimaKakuraiIsobeUedaSawa97,KitaokaKobayashiKodaWakabayashiNiinoYamakageTaguchiAmayaYamauraTakanoHiranoKanno98,TokuraNagaosa00},
where the electron spin is coupled to orbital degrees of freedom, have
been described theoretically by models with SU(4)
symmetry~\cite{LiMaShiZhang98,FrischmuthMilaTroyer99,MilaFrischmuthDeppelerTroyer99,AzariaGogolinLecheminantNerseyan99,BosscheAzariaLecheminantMila01}.
Furthermore, there has been a growing interest in the SU($n$)
generalization of the Hubbard
model~\cite{HonerkampHofstetter04,AssarafAzariaBoulatCaffarelLecheminant04}
motivated by possible experimental realizations in systems of
ultracold
atoms~\cite{AbrahamMcAlexanderGertonHuletCoteDalgarno97,RegalJin03}.

\begin{figure}[t]
\includegraphics[scale=0.20]{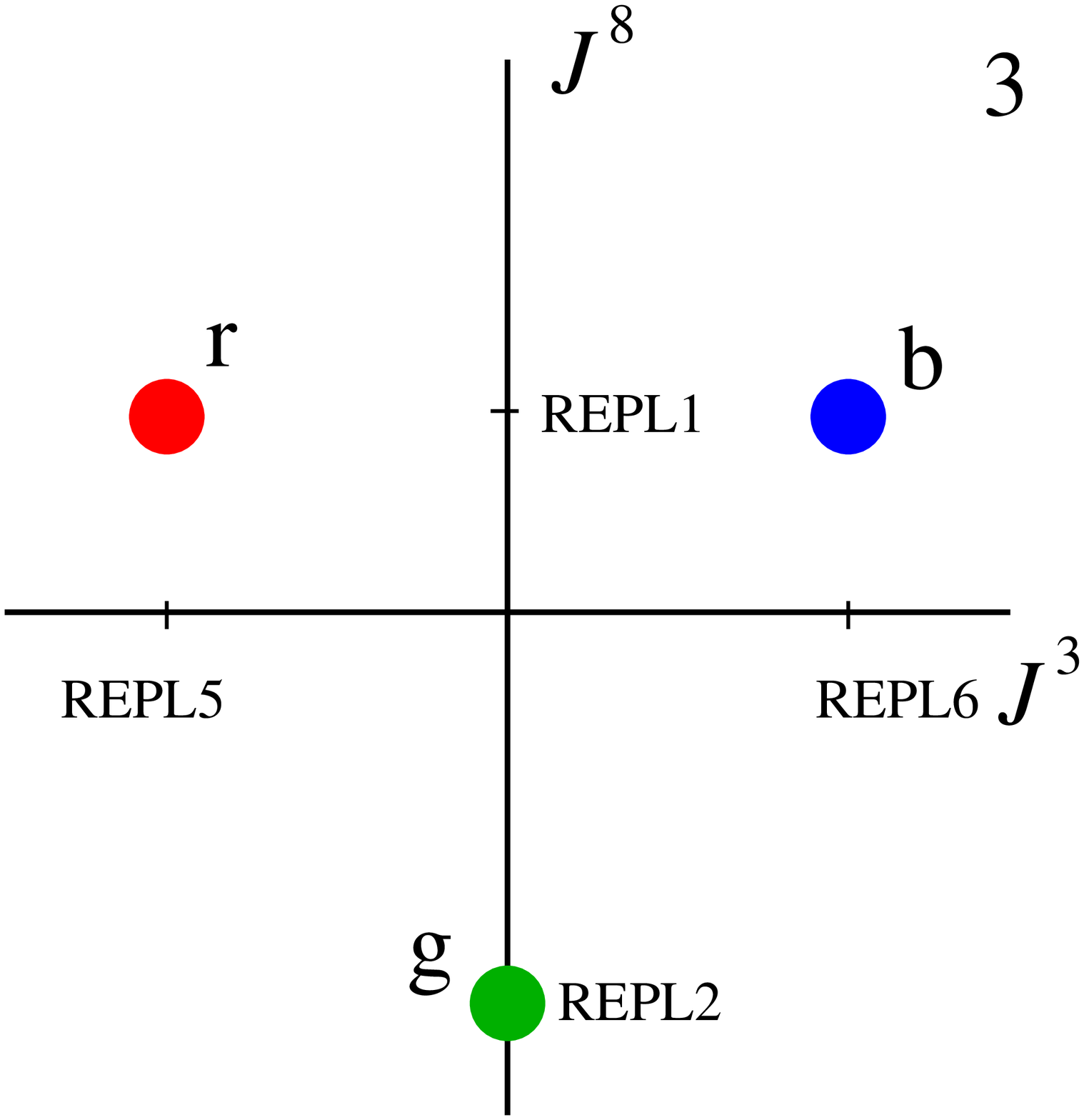}\hspace{10mm}
\includegraphics[scale=0.20]{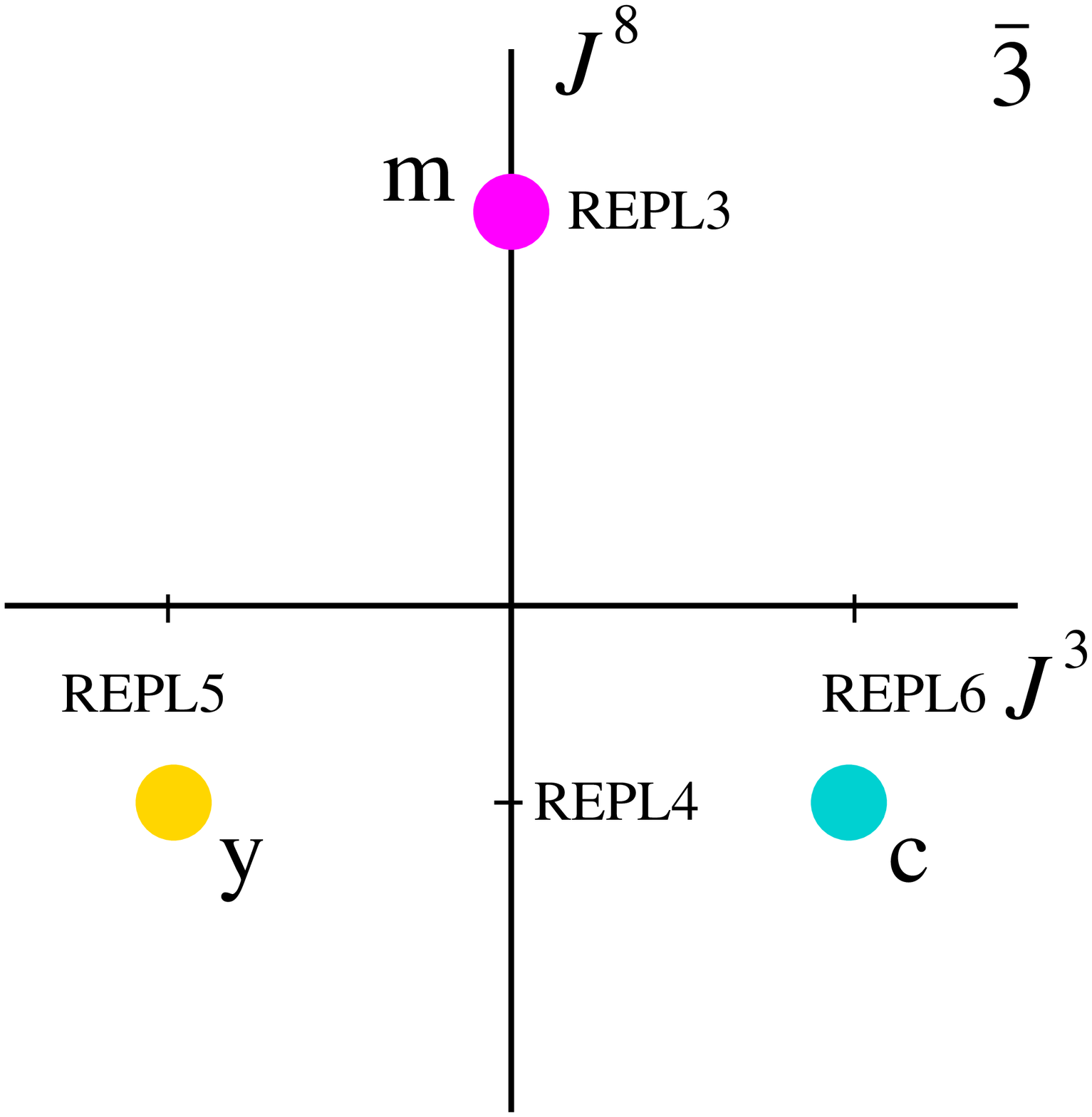}
\caption{(Color online) Weight diagrams of the SU(3) representations 
  $3$ and $\bar{3}$. $J^3$ and $J^8$ denote the diagonal
  generators~\cite{Georgi82}.}
\label{fig:reps}
\end{figure}

In a previous paper~\cite{SchurichtGreiter05colepl}, we have used the
SU($n$) Haldane--Shastry model to make the case that the elementary
excitations of SU($n$) spin chains transform under the representation
\emph{conjugate} to the representation of the SU($n$) spins on the
chain, and only exist in one n$^\text{th}$ of the Brillouin zone.  In
this article, we will present for one thing the detailed calculations
underlying our line of argumentation
in~\cite{SchurichtGreiter05colepl}.  To this end, we focus on the
SU(3) Haldane--Shastry model
(HSM)~\cite{Haldane88,Shastry88,Haldane91prl1,HaldaneHaTalstraBernardPasquier92,Kawakami92prb1,Kawakami92prb2,HaHaldane92,HaHaldane93},
which serves as a paradigm for not only the SU(3) spin chain, but, as
we shall see, also illustrates some very general properties of SU($n$)
chains.  In particular, we derive the quantum numbers of the
elementary, fractionally quantized excitations, the analogs of the
spinon excitations for SU(2), which we call colorons for SU(3).  As
already mentioned, the key result is that these excitations transform
under the SU($n$) representation conjugate to the representation of
the original SU($n$) spins localized at the sites of the chain.  In
the case of SU(3), if a basis for the original spins is spanned by the
colors blue, red, and green, a basis for the coloron excitations is
given by the complementary colors yellow, cyan, and magenta (see
Fig.~\ref{fig:reps}).  This result is meaningless for SU(2), as the
representations of SU(2) are self-conjugate, but significant in all
other instances of fractional quantization in SU($n$) chains and
possible higher dimensional liquids, regardless of model specifics.
In our analysis, however, we will primarily focus on the SU(3) HSM, as
this is the simplest model in which this general result can be
illustrated through exact calculations, and only afterwards generalize
our results to the case of SU($n$).
%
The other significant advancement we report here is the construction
of the exact, color-polarized two-coloron eigenstates.  We obtain
those through derivation and solution of a Sutherland-type equation
which is similar to the case of two spinons in the SU(2) HSM.  Our
explicit calculation implies that the colorons are, just like the
spinons, non-interacting or free, but that the spacings in the
difference of the individual coloron momenta $p_n$ and $p_m$ with
$p_n\ge p_m$ are given by
$p_n-p_m=\frac{2\pi}{N}\!\left(\frac{2}{3}+\text{integer}\right)$.  We
will argue that this spacing is a direct manifestation of the
fractional statistics~\cite{Wilczek1990,Haldane91prl2} of the colorons
with statistical exclusion parameter $g=2/3$.  This value is
consistent with what we find by naive state counting.
We then proceed by calculating the SU(3) spin currents, \ie the
eigenvalues of the diagonal components of the rapidity operator
$\Lambda^3$ and $\Lambda^8$, for the one- and two-coloron states.
Thereafter, we interpret our results within the framework of the
asymptotic Bethe Ansatz.  Finally, we generalize some of our
conclusions to the case of SU($n$).  Many details of our calculations
have been exported, largely in the form of theorems and their proofs,
into an extensive list of appendices.

\section{SU(3) Haldane--Shastry model}

The SU(3) $1/r^2$ or Haldane--Shastry model~\cite{Kawakami92prb1,HaHaldane92}
is most conveniently formulated by embedding the one-dimensional chain with
periodic boundary conditions into the complex plane by mapping it onto the
unit circle with the SU(3) spins located at complex positions
$\eta_\alpha=\exp\!\left(i\frac{2\pi}{N}\alpha\right)$, where $N$ denotes the
number of sites and $\alpha=1,\ldots,N$.  The Hamiltonian is given by
\begin{equation}
  \label{eq:su3ham}
  H_{\text{SU(3)}}
  =\left(\frac{2\pi}{N}\right)^{\!\!2}
  \sum^N_{\alpha<\beta}\frac{\bs{J}_{\alpha}\!\cdot\!
    \bs{J}_{\beta}}{\vert \eta_{\alpha}-\eta_{\beta}\vert^2},
\end{equation}
where $\bs{J}_{\alpha}=\frac{1}{2}\sum_{\sigma\tau} c_{\alpha\sigma}^{\dagger}
\bs{\lambda}_{\sigma\tau} c_{\alpha\tau}^{\phantom{\dagger}}$ is the
8-dimensional SU(3) spin vector, $\bs{\lambda}$ a vector consisting of the
eight Gell-Mann matrices (see App.~\ref{app:conventions}), and $\sigma$ and
$\tau$ are SU(3) spin or color indices, which take the values blue (b), red
(r), or green (g).  For all practical purposes, it is convenient to express
$H_{\text{SU(3)}}$ directly in terms of colorflip operators
$e_{\alpha}^{\sigma\tau}\equiv c_{\alpha\sigma}^{\dagger}
c_{\alpha\tau}^{\phantom{\dagger}}$:
\begin{equation}
  \begin{split}
  H_{\text{SU(3)}}
  &=\frac{2\pi^2}{N^2}  
  \sum^N_{\alpha<\beta}\sum_{\sigma\tau}^3\,
  \frac{\:\!e_\alpha^{\sigma\tau}\,e_\beta^{\tau\sigma}-\frac{1}{27}\:\!}
  {\vert \eta_{\alpha}-\eta_{\beta}\vert^2}\\
  &=\frac{2\pi^2}{N^2}  
  \sum^N_{\alpha<\beta}\sum_{\sigma\tau}^3\,
  \frac{e_\alpha^{\sigma\tau}\,e_\beta^{\tau\sigma}}
  {\vert \eta_{\alpha}-\eta_{\beta}\vert^2}
  -\frac{\pi^2}{N^2}\frac{N(N^2-1)}{36},
  \end{split}
  \label{eq:su3hamiltonian}
\end{equation}
where the color sum includes terms with $\sigma=\tau$. 
\psfrag{REPL1}{$\scriptstyle \frac{1}{2\sqrt{3}}$}
\psfrag{REPL2}{$\scriptstyle \frac{-1}{\sqrt{3}}$}
\psfrag{REPL3}{$\scriptstyle \frac{1}{\sqrt{3}}$}
\psfrag{REPL4}{$\scriptstyle \frac{-1}{2\sqrt{3}}$}
\psfrag{REPL5}{$\scriptstyle -\frac{1}{2}$}
\psfrag{REPL6}{$\scriptstyle \frac{1}{2}$}

The model is fully integrable even for a finite number of sites; the algebra
of infinitely many conserved quantities is
generated by the total SU(3) spin
\begin{equation}
  \label{eq:jtot}
  \comm{H_{\text{SU(3)}}}{\bs{J}}=0,
  \quad\bs{J}=\sum_{\alpha=1}^N \bs{J}_\alpha,
\end{equation}
and rapidity operators
\begin{equation}
  \label{eq:lambda}
  \comm{H_{\text{SU(3)}}}{\Lambda^a}=0,\quad
  {\Lambda}^a=\frac{1}{2}\sum_{\alpha\neq\beta}^N\,
  \frac{\eta_\alpha + \eta_\beta}{\eta_\alpha - \eta_\beta}\,
  f^{abc}{J}_\alpha^b {J}_\beta^c,
\end{equation}
where $a,b,c=1,\ldots,8$ and $f^{abc}$ are the structure constants of SU(3)
defined through $\comm{\lambda^a}{\lambda^b}=2f^{abc}\lambda^c$.
The total SU(3) spin and rapidity operators do not commute mutually,
\begin{equation}
  \label{eq:vectorlambda}
  \comm{J^a}{\Lambda^b}=f^{abc}\Lambda^c,
\end{equation}
and generate the infinite dimensional Yangian algebra
$Y(\text{sl}_3)$~\cite{HaldaneHaTalstraBernardPasquier92,ChariPressley98}.

\section{Ground state}

\begin{figure}[t]
\includegraphics[scale=0.20]{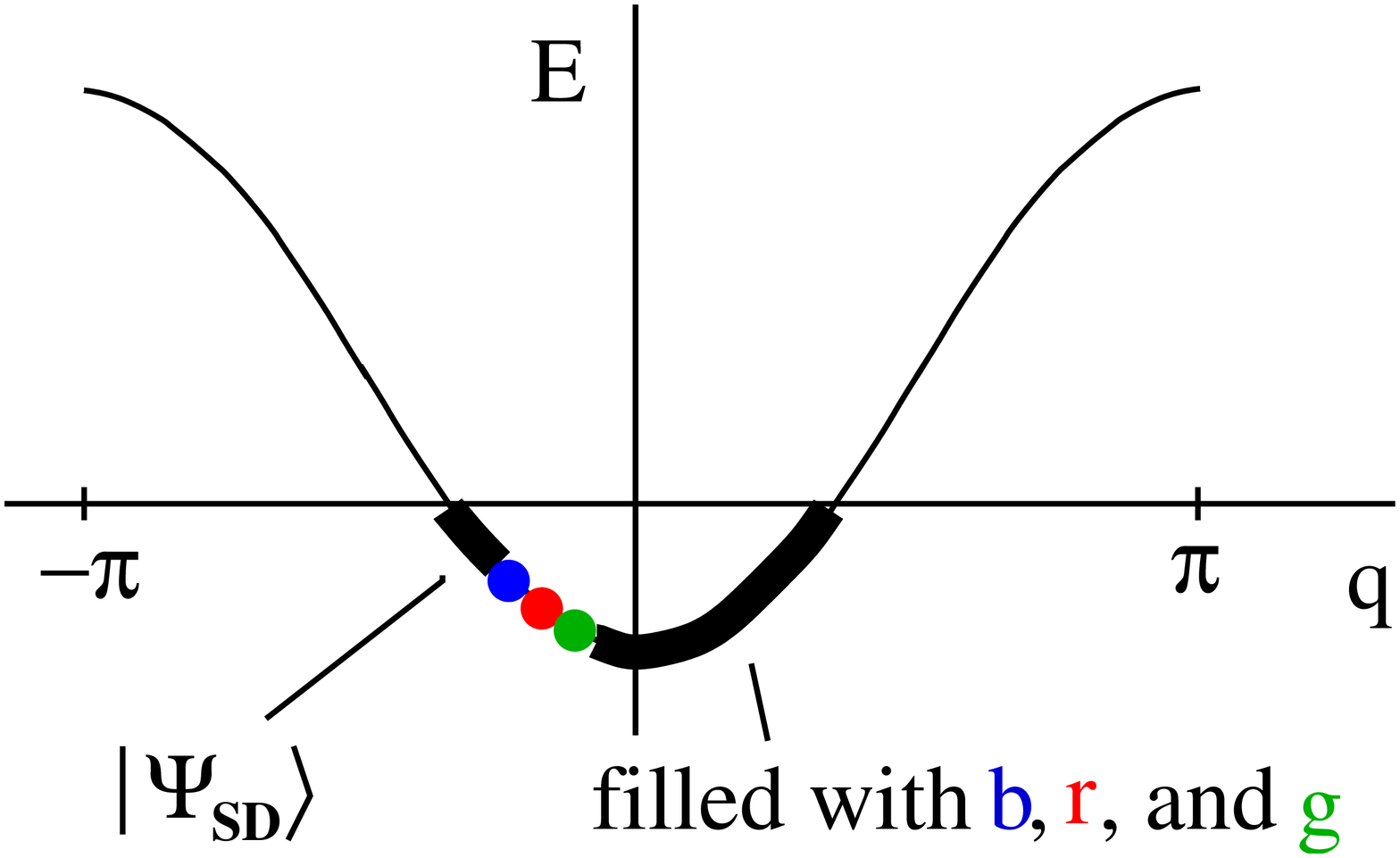}\hspace{10mm}
\includegraphics[scale=0.19]{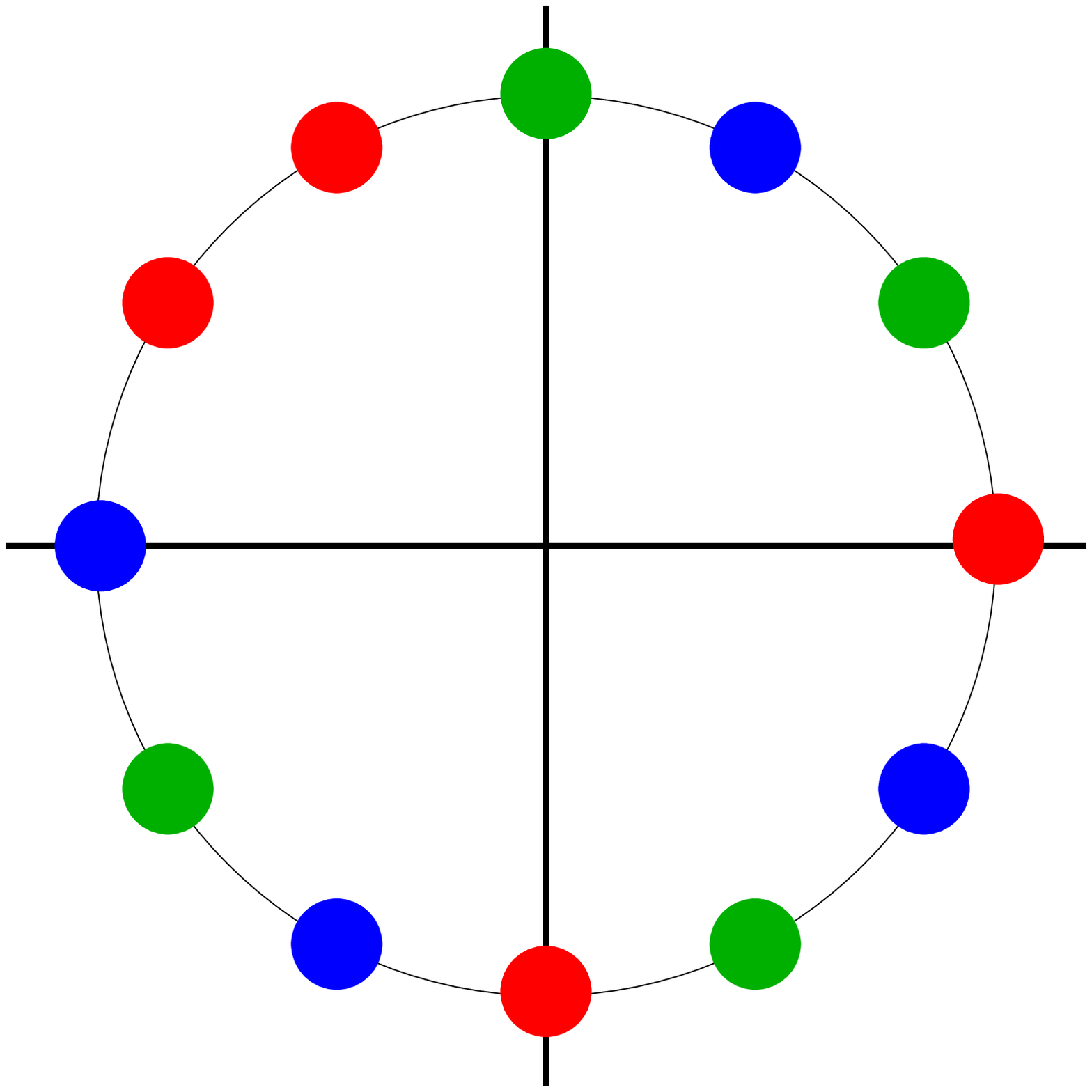}
\caption{(Color online) a) Total antisymmetric $N$-particle state. 
b) Typical configuration in $\ket{\Psi_0}$.}
\label{fig:sd}
\end{figure}

The ground state of $H_{\text{SU(3)}}$ for $N=3M$ ($M$ integer) is
most easily formulated by Gutzwiller projection of a filled band (or
Slater determinant (SD) state) containing a total of $N$ SU(3)
particles obeying Fermi statistics (see Fig.~\ref{fig:sd}a)
\begin{equation}
  \ket{\Psi_0}=P_ {\text{G}}\prod_{|q|\le q_{\text{F}}}
  c_{q\text{b}}^\dagger\,c_{q\text{r}}^\dagger\,c_{q\text{g}}^\dagger
  \ket{0}
  \equiv P_{\text{G}}\ket{\Psi_{\text{SD}}^N}\!.
  \label{eq:su3-nnhgroundstate}
\end{equation}
The Gutzwiller projector 
\begin{equation}
  \label{eq:gwproj}
  P_{\text{G}}=
  \prod_{\alpha=1}^N\bigl(n_\alpha-2\bigr)\bigl(n_\alpha-3\bigr)
\end{equation}
with $n_\alpha\equiv
c_{\alpha\b}^{\dagger}c_{\alpha\b}^{\phantom{\dagger}}+
c_{\alpha\r}^{\dagger}c_{\alpha\r}^{\phantom{\dagger}}+
c_{\alpha\g}^{\dagger}c_{\alpha\g}^{\phantom{\dagger}}$ eliminates
configurations with more than one particle on any site, and, as the
total number of particles equals the total number of sites, thereby
effectively enforces single occupancy on all sites.  As
$\ket{\Psi_{\text{SD}}^N}$ is an SU(3) singlet by construction and
$P_{\text{G}}$ commutes with SU(3) rotations, $\ket{\Psi_0}$ is an
SU(3) singlet as well.

If one interprets the state $\ket{0_\g}\equiv\prod_{\alpha=1}^N
c_{\alpha\g}^\dagger\vac$ as a reference state and the colorflip operators
$e^{\b\g}$ and $e^{\r\g}$ as ``particle creation operators'', the
ground state (\ref{eq:su3-nnhgroundstate}) can be rewritten
as~\cite{Kawakami92prb2,HaHaldane92}
\begin{equation}
  \ket{\Psi_0}=\sum_{\{z_i,w_k\}}\Psi_0[z_i;w_k]\;
  e_{z_1}^{\b\g}\ldots e_{z_M}^{\b\g}
  e_{w_1}^{\r\g}\ldots e_{w_M}^{\r\g}\ket{0_\g}\!,
\end{equation}
where the sum extends over all possible ways to distribute the
positions of the blue particles $z_1,\ldots,z_M$ and red particles
$w_1,\ldots,w_M$ over the $N$ sites (see Fig.~\ref{fig:sd}b).  The
ground state wave function is given by
\begin{equation}
    \Psi_0[z_i;w_k]\equiv
    \prod^{M_1}_{i<j}(z_i-z_j)^2\prod^{M_2}_{k<l}(w_k-w_l)^2
    \prod_{i=1}^{M_1}\prod_{k=1}^{M_2}(z_i-w_k)
    \prod_{i=1}^{M_1}z_i\prod_{k=1}^{M_2}w_k
  \label{eq:su3-definitionpsi0}
\end{equation}
with $M_1=M_2=M$, as derived in App.~\ref{app:gwstate}. The ground
state energy is
\begin{equation}
  E_0=-\frac{\pi^2}{18}\left(N+\frac{7}{N}\right)\!.
  \label{eq:su3-gsenergyM=N/3}
\end{equation}
The total momentum, as defined through $e^{ip}=\Psi_0[\eta_1
z_i,\eta_1 w_k]/\Psi_0[z_i,w_k]$ with $\eta_1=\exp(i\frac{2\pi}{N})$,
is $p=0$ regardless of $M$. (This is only true for SU($n$) with $n$
odd.)

We will now prove by explicit calculation that
(\ref{eq:su3-definitionpsi0}) is an exact eigenstate of the
Hamiltonian (\ref{eq:su3hamiltonian}). Parts of this calculation will
be used in Sec.~\ref{sec:coloronwf} to prove the exactness of our
proposed wave function for the coloron excitation. 

To evaluate the action of $H_{\text{SU(3)}}$ on
(\ref{eq:su3-definitionpsi0}) we first replace
$e_\alpha^{\g\g}e_\beta^{\g\g}$ in (\ref{eq:su3hamiltonian}) by
$(1-e_\alpha^{\b\b}-e_\alpha^{\r\r})(1-e_\beta^{\b\b}-e_\beta^{\r\r})$,
\begin{eqnarray}
H_\text{SU(3)}&=&\phantom{+}\frac{2\pi^2}{N^2}
\sum^N_{\alpha\neq\beta}\frac{1}{\vert\eta_\alpha-\eta_\beta\vert^2}
\left(e_\alpha^{\b\g}e_\beta^{\g\b}+
e_\alpha^{\r\g}e_\beta^{\g\r}+e_\alpha^{\b\r}e_\beta^{\r\b}\right)\nonumber\\
& &+\frac{2\pi^2}{N^2}\sum^N_{\alpha\neq\beta}
\frac{1}{\vert\eta_\alpha -\eta_\beta\vert^2}
\left(e_\alpha^{\b\b}e_\beta^{\b\b}+
e_\alpha^{\r\r}e_\beta^{\r\r}+e_\alpha^{\b\b}e_\beta^{\r\r}\right)
\label{eq:appsu3-hamiltonianforgs}\\
& &-\frac{2\pi^2}{N^2}\sum^N_{\alpha\neq\beta}
\frac{1}{\vert\eta_\alpha -\eta_\beta\vert^2}
\left(e_\alpha^{\b\b}+e_\alpha^{\r\r}\right)
+\frac{2}{3}\frac{\pi^2}{N^2}\sum^N_{\alpha\neq\beta}
\frac{1}{\vert\eta_\alpha -\eta_\beta\vert^2}\;,\nonumber
\end{eqnarray}
and then evaluate each term separately.

The first term $[e_\alpha^{\b\g}e_\beta^{\g\b}\Psi_0][z_i;w_k]$,
which vanishes unless one of the $z_i$'s is equal to $\eta_\alpha$,
yields through Taylor expansion
\begin{eqnarray}
\left[\sum_{\alpha\neq\beta}^N
\frac{e_\alpha^{\b\g}e_\beta^{\g\b}}{\vert\eta_\alpha-\eta_\beta\vert^2}
\Psi_0\right]\!\![z_i;w_k]&=&
\sum_{i=1}^{M_1}\sum_{\beta\neq i}^N
\frac{\eta_\beta}{\vert z_i-\eta_\beta\vert^2}
\frac{\Psi_0[\dots,z_{i-1},\eta_\beta,z_{i+1},\dots;w_k]}{\eta_\beta}
\nonumber\\
&=&\sum_{i=1}^{M_1}\sum_{m=0}^{N-1}
\frac{A_mz_i^{m+1}}{m!}
\frac{\partial^m}{\partial z_i^m}\frac{\Psi_0}{z_i}\nonumber\\ 
&=&\frac{M_1}{12}(N^2+8M_1^2-6M_1(N+1)+3)\,\Psi_0\label{eq:1331}\\
& &-\frac{N-3}{2}\sum_{i=1}^{M_1}\sum_{k=1}^{M_2}
\frac{z_i}{z_i-w_k}\Psi_0
+\sum_{i\neq j}^{M_1}\frac{z_i^2}{(z_i-z_j)^2}\Psi_0\label{eq:1331a}\\ 
& &+2\sum_{i\neq j}^{M_1}\sum_{k=1}^{M_2}
\frac{z_i^2}{(z_i-z_j)(z_i-w_k)}\Psi_0\label{eq:1331b}\\
& &+\frac{1}{2}\sum_{i=1}^{M_1}\sum_{k\neq l}^{M_2}
\frac{z_i^2}{(z_i-w_k)(z_i-w_l)}\Psi_0,\label{eq:1331c}
\end{eqnarray}
where we have used $\text{deg}_{z_i}\Psi_0[z_i;w_k]=N-1$ and defined
$A_m\equiv-\sum_{\alpha=1}^{N-1} \eta_\alpha^2 (\eta_\alpha -1)^{m-2}$.
Evaluation of the latter yields $A_0=(N-1)(N-5)/12$, $A_1=-(N-3)/2$, $A_2=1$,
and $A_m=0$ for $2<m\le N-1$ (see App.~\ref{app:formulas}).  
Furthermore, we have used
\begin{equation}
\frac{x^2}{(x-y)(x-z)}+\frac{y^2}{(y-x)(y-z)}+\frac{z^2}{(z-x)(z-y)}=1,
\quad x,y,z\in\mathbb{C}.
\label{eq:appsu3-threezformula}
\end{equation}
The second term $[e_\alpha^{\r\g}e_\beta^{\g\r}\Psi_0][z_i;w_k]$ can
be treated in the same way and yields together with the first term in
(\ref{eq:1331a})
\begin{displaymath}
-\frac{N-3}{2}\sum_{i=1}^{M_1}\sum_{k=1}^{M_2}\frac{z_i}{z_i-w_k}
+\frac{N-3}{2}\sum_{i=1}^{M_1}\sum_{k=1}^{M_2}\frac{w_k}{z_i-w_k}=
-\frac{N-3}{2}M_1M_2,
\end{displaymath} 
and with one part of (\ref{eq:1331b}) and (\ref{eq:1331c})
\begin{displaymath}
\sum_{\substack{i,j=1\\i\neq j}}^{M_1}\sum_{k=1}^{M_2}\left(
\frac{z_i^2}{(z_i-z_j)(z_i-w_k)}+\frac{1}{2}\frac{w_k^2}{(z_i-w_k)(z_j-w_k)}
\right)=\frac{1}{2}M_1(M_1-1)M_2,
\end{displaymath}
as well as similar expressions for $z_i\leftrightarrow w_k$.

For the third term of (\ref{eq:appsu3-hamiltonianforgs}) we obtain
\begin{eqnarray}
\left[\sum_{\alpha\neq\beta}^N
\frac{e_\alpha^{\b\r}e_\beta^{\r\b}}{\vert\eta_\alpha-\eta_\beta\vert^2}
\Psi_0\right]&&\!\!\!\!\!\!\!\!\!\!\![z_i;w_k]=
\sum_{i=1}^{M_1}\sum_{k=1}^{M_2}\frac{z_iw_k}{(z_i-w_k)^2}
\prod^{M_1}_{j\neq i}\left(1+\frac{z_i-w_k}{z_j-z_i}\right)
\prod^{M_2}_{l\neq k}\left(1-\frac{z_i-w_k}{w_l-w_k}\right)
\Psi_0\nonumber\\
&=&\!\!\!\phantom{-}\sum_{i=1}^{M_1}\sum_{k=1}^{M_2}
\frac{z_iw_k}{(z_i-w_k)^2}\Psi_0\label{eq-appsu3-1221first}\\
& &\!\!\!-\sum_{i\neq j}^{M_1}\sum_{k=1}^{M_2}
\frac{z_iw_k}{(z_i-z_j)(z_i-w_k)}\Psi_0-
\sum_{i=1}^{M_1}\sum_{k\neq l}^{M_2}
\frac{z_iw_k}{(w_k-z_i)(w_k-w_l)}\Psi_0\label{eq:appsu3-1221second}\\
& &\!\!\!+\sum_{i=1}^{M_1}\sum_{k=1}^{M_2}\sum_{m=2}^{M_1-1}\frac{1}{m!}
\sum_{\{a_j\}}\frac{z_iw_k(z_i-w_k)^{m-2}}{(z_{a_1}-z_i)\cdots
(z_{a_m}-z_i)}\Psi_0\label{eq:appsu3-1221third}\\
& &\!\!\!+\sum_{i=1}^{M_1}\sum_{k=1}^{M_2}\sum_{n=2}^{M_2-1}\frac{(-1)^n}{n!}
\sum_{\{b_l\}}\frac{z_iw_k(z_i-w_k)^{n-2}}{(w_{b_1}-w_k)\cdots
(w_{b_n}-w_k)}\Psi_0\label{eq:appsu3-1221forth}\\
& &\!\!\!+\sum_{i=1}^{M_1}\sum_{k=1}^{M_2}\sum_{m=1}^{M_1-1}\sum_{n=1}^{M_2-1}
\frac{(-1)^n}{m!n!}\nonumber\\
& &\!\!\!\quad\;\sum_{\{a_j\}\{b_l\}}
\frac{z_iw_k(z_i-w_k)^{m+n-2}}{(z_{a_1}-z_i)\cdots(z_{a_m}-z_i)
(w_{b_1}-w_k)\cdots(w_{b_n}-w_k)}\Psi_0\label{eq:appsu3-1221fifth},
\end{eqnarray}
where $\{a_j\}$ ($\{b_l\}$) is a set of integers between 1 and $M_1$ ($M_2$).
The summations run over all possible ways to distribute the $z_{a_j}$
($w_{b_l}$) over the blue (red) coordinates, where $z_i$ ($w_k$) is excluded.
The two terms (\ref{eq:appsu3-1221third}) and (\ref{eq:appsu3-1221forth}) 
vanish due to 
\begin{theo}\label{theo:gstheorem}
  Let $M\ge 3$, $z\in\mathbb{C}$, and $z_1,\ldots, z_M\in\mathbb{C}$ distinct.
  Then,
\begin{equation}
\sum_{i=1}^M\frac{z_i(z_i-z)^{M-3}}{\prod_{j\neq i}^M(z_j-z_i)}=0.
\label{eq:theo1}
\end{equation}
\end{theo}
A proof of Theorem \ref{theo:gstheorem} is given in
App.~\ref{app:gstheorem}.  The last term (\ref{eq:appsu3-1221fifth})
can be simplified using a theorem due to Ha and
Haldane~\cite{HaHaldane92}:
\begin{theo}\label{theo:appsu3-hahaldane}
  Let $\{a_j\}$ be a set of distinct integers between $1$ and $M_1$, and
  $\{b_l\}$ a set of distinct integers between $1$ and $M_2$. Then,
\begin{displaymath}
\begin{split}
\sum_{i=1}^{M_1}\sum_{k=1}^{M_2}\sum_{m=1}^{M_1-1}\sum_{n=1}^{M_2-1}
\sum_{\{a_j\}\{b_l\}}\frac{(-1)^n}{m!n!}&
\frac{z_iw_k(z_i-w_k)^{m+n-2}}{(z_{a_1}-z_i)\cdots(z_{a_n}-z_i)
(w_{b_1}-w_k)\cdots(w_{b_m}-w_k)}\\
&=-\sum_{m=1}^{\mathrm{min}(M_1,M_2)}\!\!\!\!(M_1-m)(M_2-m).
\end{split}
\end{displaymath}
\end{theo}
Furthermore, the two terms in line (\ref{eq:appsu3-1221second}),
together with the remainder of (\ref{eq:1331b}) and the corresponding
expression from the second term of the Hamiltonian, can be simplified
to yield $M_1M_2(M_1+M_2-2)\Psi_0/2$.

Finally, the diagonal terms of (\ref{eq:appsu3-hamiltonianforgs})
cancel the remainder of (\ref{eq:1331a}) as well as
(\ref{eq-appsu3-1221first}) and yield the additional constant
\begin{displaymath}
\frac{2\pi^2}{N^2}\left[\frac{1}{2}\Bigl(M_1(M_1-1)+M_2(M_2-1)\Bigr)
+\frac{N^2-1}{12}\Bigl(\frac{N}{3}-M_1-M_2\Bigr)\right]\!.
\end{displaymath}
When collecting all terms evaluated above and setting $M_1=M_2=N/3$ we
obtain the ground state energy (\ref{eq:su3-gsenergyM=N/3}).

\section{One-coloron states}

\subsection{Trial wave functions}

We now turn to the heart of our analysis, the elementary and fractionally
quantized excitations, which we call colorons.  In principle, there are two
possible, non-equivalent constructions for localized excitations starting from
(\ref{eq:su3-nnhgroundstate}).  We may either create a particle with color
$\sigma$ on a chain with $N=3M+1$ sites before Gutzwiller projection,
\begin{equation}
  \ket{\Psi_{\gamma\sigma}^{\text{c}}}=
  P_{\text{G}}\;\!c_{\gamma\sigma}^\dagger\!\ket{\Psi_{\text{SD}}^{N-1}}\!,
  \label{eq:loctwocolorons}
\end{equation}
or annihilate a particle with color $\sigma$ on a chain with $N=3M-1$:
\begin{equation}
  \ket{\Psi_{\gamma\bar\sigma}^\text{a}}=P_{\text{G}}\;\!
  c_{\gamma\sigma}^{\phantom{\dagger}}\!\ket{\Psi_{\text{SD}}^{N+1}}\!.
  \label{eq:loccoloron}
\end{equation}
In both cases, $c_{\gamma\sigma}^\dagger$ or
$c_{\gamma\sigma}^{\phantom{\dagger}}$ creates an inhomogeneity in color and
charge before projection.  The projection once again enforces one particle per
site and thereby removes the charge inhomogeneity, while it commutes with
color and thereby preserves the color inhomogeneity.  Consequently, the trial
states $\ket{\Psi_{\gamma\sigma}^{\text{c}}}$ and
$\ket{\Psi_{\gamma\bar\sigma}^\text{a}}$ describe localized ``excitations'' of
color $\sigma$ or complementary color $\bar\sigma$, respectively, at lattice
site $\eta_\gamma$.  Since $H_{\text{SU(3)}}$ is translationally
invariant, we of course expect neither of them, but only momentum eigenstates
constructed from them via
\begin{equation}
  \ket{\Psi_n}
  \equiv\frac{1}{N}\sum_{\gamma=1}^N e^{-i\frac{2\pi}{N}\gamma n}
  \ket{\Psi_\gamma}
  =\frac{1}{N}\sum_{\gamma=1}^N (\bar\eta_\gamma)^n \ket{\Psi_\gamma},
  \label{eq:ncoloron}
\end{equation}
where $n$ is a momentum quantum number, to be eigenstates.

The important thing to realize now is that only 
(\ref{eq:loctwocolorons}) {or} (\ref{eq:loccoloron}), but not both, can
describe a valid excitation.  To see this, let us assume $\sigma=\b$ for ease
in presentation, and note that (\ref{eq:loctwocolorons}) is apart from a
normalization factor equivalent to
\begin{equation}
  \ket{\Psi_{\gamma\b}^{\text{c}}}=
   P_{\text{G}}\;\!c_{\gamma\r}^{\phantom{\dagger}}
  c_{\gamma\g}^{\phantom{\dagger}}\ket{\Psi_{\text{SD}}^{N+2}},
  \label{eq:loctwocolorons2}
\end{equation}
\ie creation of a blue particle before projection is tantamount to
annihilation of both a red and a green particle at site $\eta_\gamma$.
If momentum eigenstates constructed from
$\ket{\Psi_{\gamma\bar\sigma}^\text{a}}$ via (\ref{eq:ncoloron}) were
energy eigenstates, the anti-red (cyan) and anti-green (magenta)
coloron excitations in (\ref{eq:loctwocolorons2}) would individually
seek to be momentum eigenstates, which implies that a trial wave
function forcing them to sit on the same site would not be an energy
eigenstate.  The same argument can be made the other way round.

\subsection{One-coloron wave functions}\label{sec:coloronwf}

\begin{figure}[t]
 \setlength{\unitlength}{10pt}
 \begin{picture}(16,7)(-1,0)
 \put(0,1){\line(1,0){13.5}}
 \put(0,1){\line(0,1){5}}
 \qbezier[2000](4,1)(6,8)(8,1)
 \put(4,1){\makebox(0,0)[t]{\rule{0.3pt}{4pt}}}
 \put(8,1){\makebox(0,0)[t]{\rule{0.3pt}{4pt}}}
 \put(12,1){\makebox(0,0)[t]{\rule{0.3pt}{4pt}}}
 \put(0,0){\makebox(0,0){\small $0$}}
 \put(4,0){\makebox(0,0){\small $\frac{2\pi}{3}$}}
 \put(8,0){\makebox(0,0){\small $\frac{4\pi}{3}$}}
 \put(12,0){\makebox(0,0){\small $2\pi$}}
 \put(13.7,0.4){\makebox(0,0){\small $p$}}
 \put(-1.3,6){\makebox(0,0){\small $\epsilon(p)$}}
 \put(6,1){\makebox(0,0){\rule{40pt}{3pt}}}
 \put(6,1.5){\line(1,1){3}}
 \put(12,5.3){\makebox(0,0){\small allowed momenta}}
 \end{picture}
 \caption{One-coloron dispersion relation.}
 \label{fig:dis}
 \end{figure}
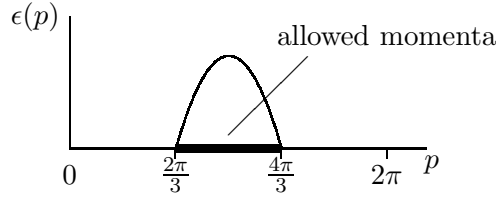

In the following, we show by explicit calculation that momentum
eigenstates constructed from $\ket{\Psi_{\gamma\bar\sigma}^\text{a}}$
via (\ref{eq:ncoloron}) are exact eigenstates of $H_{\text{SU(3)}}$.
For simplicity, we choose $\bar\sigma=\bar\b=\y$ (an anti-blue or
yellow coloron) and express
$\bigl|\Psi_{\gamma\bar\b}^\text{a}\bigr\rangle$ through the
corresponding wave function
\begin{equation}
  \Psi_\gamma[z_i;w_k]=
  \prod_{i=1}^{M_1} (\eta_\gamma-z_i)\;\Psi_0[z_i;w_k]\equiv\psi_\gamma\Psi_0,
  \label{eq:localizedcoloronwf}
\end{equation}
with $\Psi_0$ given by (\ref{eq:su3-definitionpsi0}) and
$M_1=(N-2)/3$, $M_2=(N+1)/3$.  The momentum eigenstate
$\bigl|\Psi_{n\bar\b}^\text{a}\bigr\rangle$ is then given by
\begin{equation}
  \Psi_n[z_i;w_k]=
  \frac{1}{N}\sum_{\gamma=1}^N(\bar{\eta}_\gamma)^n\;\Psi_\gamma[z_i;w_k]
  \label{eq:wfmomentumcoloron}
\end{equation}
(where $n$ is shifted with respect to $n$ in (\ref{eq:ncoloron}) by a
constant depending on which momenta are occupied in the Slater
determinant state). The momentum of (\ref{eq:wfmomentumcoloron}) is 
\begin{equation}
  p=\frac{4\pi}{3} -\frac{2\pi}{N}\left(n+\frac{1}{3}\right)\!,
  \quad 0\le n\le M_1.
\label{eq:coloronmomentum}
\end{equation}
The energy is given by
\begin{equation}
\begin{split}
E_n&=-\frac{\pi^2}{18}\left(N+\frac{1}{N}+\frac{2}{N^2}\right)\,+\,
\frac{3\pi^2}{N^2}(M_1-n)n\\[2mm]
&=E_0+\frac{2}{9}\frac{\pi^2}{N^2}\,+\,\epsilon(p),
  \label{eq:colorondisperion}
\end{split}
\end{equation} 
where we have defined the one-coloron dispersion relation (see
Fig.~\ref{fig:dis})
\begin{equation}
\epsilon(p)=\frac{3}{4}\left(\frac{\pi^2}{9}-(p-\pi)^2 \right)\!.
\label{eq:singlecolorondispersion}
\end{equation}
For $M_1<n<N$, $\Psi_n$ vanishes identically.

To prove that $\Psi_n$ represents an energy eigenstate, note that all
terms of the Hamiltonian (\ref{eq:appsu3-hamiltonianforgs}) yield the
same expressions as for the wave function $\Psi_0$ except those terms
containing $e_\alpha^{\b\g}e_\beta^{\g\b}$ and
$e_\alpha^{\b\r}e_\beta^{\r\b}$. Hence, it is sufficient to
investigate those further.

To begin with, the polynomial $\psi_\gamma$ defined in
(\ref{eq:localizedcoloronwf}) can be written as $\sum_{m=0}^{M_1}
(-1)^{M_1-m}\,\eta_\gamma^m\,\mathcal{S}^{M_1}(z_1\cdots z_{M_1-m})$,
where $\mathcal{S}^{M}(z_1\cdots z_q)$ is the sum over all possible
ways to choose $q$ coordinates out of $z_1,\ldots,z_M$, \eg
$\mathcal{S}^3(z_1z_2)=z_1z_2+z_1z_3+z_2z_3$. Hence, we find in
addition to (\ref{eq:1331})--(\ref{eq:1331c}) the two terms (note that
$\mathrm{deg}_{z_i}\Psi_\gamma= \mathrm{deg}_{w_k}\Psi_\gamma=N-1$)
\begin{equation}
\Psi_0 \sum_{i=1}^{M_1}
\left(\frac{1}{2}z_i^2\frac{\partial^2}{\partial z_i^2}
+\sum_{j\neq i}^{M_1} \frac{2z_i^2}{z_i-z_j}
\frac{\partial}{\partial z_i}-\frac{N-3}{2}z_i\frac{\partial}{\partial z_i}
\right)\psi_\gamma+\Psi_0 \sum_{i=1}^{M_1}\sum_{k=1}^{M_2}\frac{z_i^2}{z_i-w_k}
\frac{\partial}{\partial z_i}\;\psi_\gamma\label{eq:appsu3-os1331third}.
\end{equation}
Fourier transformation (\ref{eq:wfmomentumcoloron}) of these yields
\begin{equation}
\Biggl[M_1\left(M_1-\frac{N-1}{2}\right)-
n\left(n-\frac{N-1}{2}\right)\Biggr]\Psi_n
-\frac{1}{N}\Psi_0\sum_{\gamma=1}^N\sum_{i=1}^{M_1}\sum_{k=1}^{M_2}
\frac{z_i^2(\bar{\eta}_\gamma)^n}{(z_i-w_k)(\eta_\gamma-z_i)}\;\psi_\gamma.
\label{eq:appsu3-firstresult1331term}
\end{equation}
Furthermore, 
\begin{eqnarray}
\left[\sum_{\alpha\neq\beta}^N
\frac{e_\alpha^{\b\r}e_\beta^{\r\b}}{\vert\eta_\alpha-\eta_\beta\vert^2}
\Psi_n\right]\!\![z_i;w_k]
&=&\sum_{i=1}^{M_1}\sum_{k=1}^{M_2}\frac{z_iw_k}{(z_i-w_k)^2}
\prod^{M_1}_{j\neq i}\frac{z_j-w_k}{z_j-z_i}
\prod^{M_2}_{l\neq k}\frac{w_l-z_i}{w_l-w_k}
\Psi_n\label{eq:appsu3-os1221first}\\
& &\hspace{-20mm}+\frac{\Psi_0}{N}\sum_{\gamma=1}^N\sum_{i=1}^{M_1}
\sum_{k=1}^{M_2}\frac{z_iw_k(\bar{\eta}_\gamma)^n}{(z_i-w_k)(\eta_\gamma-z_i)}
\prod^{M_1}_{j\neq i}\frac{z_j-w_k}{z_j-z_i}
\prod^{M_2}_{l\neq k}\frac{w_l-z_i}{w_l-w_k}
\psi_\gamma.\label{eq:appsu3-os1221second}
\end{eqnarray}
The term on the right of (\ref{eq:appsu3-os1221first}) is identical to
a corresponding term in the ground state calculation, whereas
(\ref{eq:appsu3-os1221second}) taken together with the second term in
(\ref{eq:appsu3-firstresult1331term}) gives
\begin{displaymath}
\left[-\frac{1}{2}n(n+1)+\frac{1}{2}M_1(M_1+1)\right]\Psi_n,
\end{displaymath}
where we have used
\begin{theo}\label{theo:appsu3-onespinontheorem}
  For $M_1,M_2\in\mathbb{N}$, $M_1\le M_2$,
  $z_1,\dots,z_{M_1}\in\mathbb{C}$ distinct,
  $w_1,\dots,w_{M_2}\in\mathbb{C}$ distinct, and $0\le n\le M_1$
  the following is valid:
\begin{equation}
\begin{split}
\sum_{i=1}^{M_1}\sum_{k=1}^{M_2}&\frac{z_i}{z_i-w_k}
\left[w_k\prod^{M_1}_{j\neq i}\frac{z_j-w_k}{z_j-z_i}
\prod^{M_2}_{l\neq k}\frac{w_l-z_i}{w_l-w_k}-z_i\right]
\sum^N_{\gamma=1}(\bar{\eta}_\gamma)^n
\prod^{M_1}_{j\neq i}(\eta_\gamma-z_j)\\[3mm]
&=\left[-\frac{1}{2}n(n+1)+\frac{1}{2}M_1(M_1+1)\right]
\sum^N_{\gamma=1}(\bar{\eta}_\gamma)^n\prod^{M_1}_{i=1}(\eta_\gamma-z_i).
\end{split}
\label{eq:appsu3-spinontheorem}
\end{equation}
\end{theo}
The proof of Theorem~\ref{theo:appsu3-onespinontheorem} is given in
App.~\ref{app:onecolorontheorem}.  Collecting the terms obtained
above, we find that $\Psi_n$ is an exact eigenstate of the Hamiltonian
(\ref{eq:su3ham}) with energy (\ref{eq:colorondisperion}).

\subsection{Quantum numbers of colorons}
\label{sec:quantumnumbers}

In the preceding section we have shown that the elementary coloron
excitations are constructed by annihilation of a particle of color
$\sigma$ from an overall color singlet $\ket{\Psi_{\text{SD}}^{N+1}}$
\emph{before} Gutzwiller projection.  Since they are hole-like
excitations, they transform according to the representation $\bar 3$
conjugate to the fundamental representation $3$ of the original
particles on the sites of the chain.  This can also be seen by acting
with the total SU(3) spin generators (\ref{eq:jtot}) on the wave
function (\ref{eq:localizedcoloronwf}).  Our result agrees with
results obtained for the spectrum of the SU(3)$_1$
Wess--Zumino--Witten (WZW) model by Bouwknegt and
Schoutens~\cite{BouwknegtSchoutens96,Schoutens97}.

It is straightforward to read off the quantum or exclusion
statistics~\cite{Wilczek1990,Haldane91prl2} of color-polarized
colorons (\ie colorons of the same color).  Consider a chain with
$N=3M-1$ sites and a single yellow coloron.  According to
(\ref{eq:su3-nnhgroundstate}), (\ref{eq:loccoloron}), and
(\ref{eq:ncoloron}), there are as many single particle orbitals
available to the coloron as there are blue particles in the Slater
determinant state, that is, $M$.  If we now were to create three
additional yellow colorons, the Slater determinant state would have to
contain three more particles, one of each color.  This implies there
would be one additional orbital, while the three additional colorons
would occupy three orbitals, meaning that the number of orbitals
available for our original coloron would be reduced by two.  The
statistical parameter is hence given by $g=2/3$.  The fractional
statistics manifests itself further in the exponents of the algebraic
decay of the form factor of the dynamical structure
factor~\cite{YamamotoSaigaArikawaKuramoto00prl,YamamotoSaigaArikawaKuramoto00jpsj}
as well as in the thermodynamics of the
model~\cite{KuramotoKato95,KatoKuramoto96}.  A similar exclusion
statistics exists in the conformal field theory spectrum of WZW
models~\cite{Schoutens97,BouwknegtSchoutens99}.  The exclusion
statistics among colorons of different colors as well as the state
counting for SU($n$) ($n\ge 3$) spin chains in general is highly
non-trivial and will be subject of a future
publication~\cite{manuscriptinpreparationGS}.

\section{Two-coloron states}\label{sec:twocolorons}

For $N=3M-2$, there are at least two colorons present. Two yellow
colorons localized at lattice sites $\eta_\gamma$ and $\eta_\delta$
are described by the wave function
\begin{equation}
\Psi_{\gamma\delta}[z_i;w_k]=
  \prod_{i=1}^{M_1}(\eta_\gamma-z_i)(\eta_\delta-z_i)\;\Psi_0[z_i;w_k]
  \equiv \psi_{\gamma\delta}\;\Psi_0,
\label{eq:2loccol}
\end{equation}
where $M_1=(N-4)/3$ and $M_2=(N+2)/3$.  Momentum eigenstates are
once again constructed by Fourier transformation
\begin{equation}
\Psi_{mn}[z_i;w_k]=\frac{1}{N^2}\sum_{\gamma\delta}^N
(\bar{\eta}_\gamma)^m(\bar{\eta}_\delta)^n\Psi_{\gamma\delta}[z_i;w_k],\quad
0\le n \le m \le M_1. 
\label{eq:2momcol}
\end{equation}
In analogy to the SU(2) HSM we neither expect these states for fixed
$m$ and $n$ to be energy eigenstates, nor states with different sets
of quantum numbers $(m,n)$ with the same $m+n$ (and hence the same
total momentum) to be orthogonal to each other.  Energy eigenstates,
however, can be constructed as follows.

As for the one-coloron states, application of the Hamiltonian
generates two contributions in addition to those familiar from the the
ground state calculation.  First, for the terms containing 
$e_\alpha^{\b\g}e_\beta^{\g\b}$ we obtain (\ref{eq:appsu3-os1331third})
with $\psi_\gamma$ replaced by $\psi_{\gamma\delta}$. The first term
in the resulting expression can be treated in analogy to the
two-spinon states in the SU(2)
HSM~\cite{BernevigGiulianoLaughlin01prb}.  Specifically, we find
\begin{displaymath}
\begin{split}
&\sum_{i=1}^{M_1}
\left(\frac{1}{2}z_i^2\frac{\partial^2}{\partial z_i^2}
+\sum_{j\neq i}^{M_1} \frac{2z_i^2}{z_i-z_j}
\frac{\partial}{\partial z_i}
-\frac{N-3}{2}z_i\frac{\partial}{\partial z_i}
\right)\psi_{\gamma\delta}=M_1(2M_1-N+2)\psi_{\gamma\delta}\\[2mm]
&+\left[-\eta_\gamma^2\frac{\partial^2}{\partial\eta_\gamma^2}
-\eta_\delta^2\frac{\partial^2}{\partial\eta_\delta^2}
+\frac{N-3}{2}\left(\eta_\gamma\frac{\partial}{\partial\eta_\gamma}+
\eta_\delta\frac{\partial}{\partial\eta_\delta}\right)
\right]\psi_{\gamma\delta}-\frac{1}{\eta_\gamma-\eta_\delta}\left[
\eta_\gamma^2\frac{\partial}{\partial\eta_\gamma}-
\eta_\delta^2\frac{\partial}{\partial\eta_\delta}
\right]\psi_{\gamma\delta}
\end{split}
\end{displaymath}
where we have used (\ref{eq:appsu3-threezformula}) three times. 
Applying this identity to the momentum eigenstates (\ref{eq:2momcol}),
we obtain
\begin{displaymath}
\begin{split}
-M_1(M_1+2)\Psi_{mn}+&\left[\left(\frac{3}{2}M_1+1-m\right)m+
\left(\frac{3}{2}M_1+1-n\right)n-\frac{m-n}{2}\right]\Psi_{mn}\\[2mm]
&-\sum_{\ell=1}^{\ell_\mathrm{m}}(m-n+2\ell)\Psi_{m+\ell,n-\ell},
\end{split}
\end{displaymath}
where $\ell_\mathrm{m}=\mathrm{min}(M_1-m,n)$, and we have used 
\begin{equation}
\frac{x+y}{x-y}(x^my^n-x^ny^m)=
2\sum_{\ell=0}^{m-n}x^{m-\ell}y^{n+\ell}-(x^my^n+x^ny^m),\quad m\ge n.
\label{eq:xysum}
\end{equation}
The second term of (\ref{eq:appsu3-os1331third}) applied to the states
$\Psi_{mn}$ yields
\begin{equation}
-\frac{1}{N^2}\Psi_0\sum_{\gamma\delta}^N\sum_{i=1}^{M_1}\sum_{k=1}^{M_2}
\frac{z_i^2(\bar{\eta}_\gamma)^m(\bar{\eta}_\delta)^n}{z_i-w_k}
\left(\frac{1}{\eta_\gamma-z_i}+\frac{1}{\eta_\delta-z_i}\right)
\psi_{\gamma\delta},
\label{eq:twocolaux1}
\end{equation}

The second contribution not familiar from the ground state calculation
is given by terms arising in 
\begin{eqnarray}
\left[\sum_{\alpha\neq\beta}^N
\frac{e_\alpha^{\b\r}e_\beta^{\r\b}}{\vert\eta_\alpha-\eta_\beta\vert^2}
\Psi_{mn}\right]\!\![z_i;w_k]
&=&\sum_{i=1}^{M_1}\sum_{k=1}^{M_2}\frac{z_iw_k}{(z_i-w_k)^2}
\prod^{M_1}_{j\neq i}\frac{z_j-w_k}{z_j-z_i}
\prod^{M_2}_{l\neq k}\frac{w_l-z_i}{w_l-w_k}
\Psi_{mn}\label{eq:twocolaux2}\\
& &\hspace{-25mm}+\frac{\Psi_0}{N^2}\sum_{\gamma\delta}^N\sum_{i=1}^{M_1}
\sum_{k=1}^{M_2}\frac{z_iw_k(\bar{\eta}_\gamma)^m(\bar{\eta}_\delta)^n}
{(z_i-w_k)(\eta_\gamma-z_i)}
\prod^{M_1}_{j\neq i}\frac{z_j-w_k}{z_j-z_i}
\prod^{M_2}_{l\neq k}\frac{w_l-z_i}{w_l-w_k}
\psi_{\gamma\delta}\label{eq:twocolaux3}\\
& &\hspace{-25mm}+\frac{\Psi_0}{N^2}\sum_{\gamma\delta}^N\sum_{i=1}^{M_1}
\sum_{k=1}^{M_2}\frac{z_iw_k(\bar{\eta}_\gamma)^m(\bar{\eta}_\delta)^n}
{(z_i-w_k)(\eta_\delta-z_i)}
\prod^{M_1}_{j\neq i}\frac{z_j-w_k}{z_j-z_i}
\prod^{M_2}_{l\neq k}\frac{w_l-z_i}{w_l-w_k}
\psi_{\gamma\delta}\label{eq:twocolaux4}\\
& &\hspace{-25mm}+\frac{\Psi_0}{N^2}\sum_{\gamma\delta}^N\sum_{i=1}^{M_1}
\sum_{k=1}^{M_2}\frac{z_iw_k(\bar{\eta}_\gamma)^m(\bar{\eta}_\delta)^n}
{(\eta_\gamma-z_i)(\eta_\delta-z_i)}
\prod^{M_1}_{j\neq i}\frac{z_j-w_k}{z_j-z_i}
\prod^{M_2}_{l\neq k}\frac{w_l-z_i}{w_l-w_k}
\psi_{\gamma\delta}.\label{eq:twocolaux5}
\end{eqnarray}
The first term (\ref{eq:twocolaux2}) is identical to the ground state
calculation.  The two following terms (\ref{eq:twocolaux3}) and
(\ref{eq:twocolaux4}) can be combined together with
(\ref{eq:twocolaux1}) and Theorem~\ref{theo:appsu3-onespinontheorem}
to yield
\begin{equation}
\left[M_1(M_1+1)-\frac{1}{2}m(m+1)-\frac{1}{2}n(n+1)\right]\Psi_{mn}.
\end{equation}
The last term (\ref{eq:twocolaux5}) can be simplified using
\begin{theo}\label{theo:twocolorons}
  For $M_1,M_2\in\mathbb{N}$, $M_1\le M_2-1$,
  $z_1,\dots,z_{M_1}\in\mathbb{C}$ distinct,
  $w_1,\dots,w_{M_2}\in\mathbb{C}$ distinct, and $0\le n\le m\le M_1$, and 
  $\ell_\mathrm{m}=\mathrm{min}(M_1-m,n)$ the following is valid:
\begin{equation}
\begin{split}
\sum_{i=1}^{M_1}\sum_{k=1}^{M_2}&z_iw_k
\prod^{M_1}_{j\neq i}\frac{z_j-w_k}{z_j-z_i}
\prod^{M_2}_{l\neq k}\frac{w_l-z_i}{w_l-w_k}
\sum^N_{\gamma\delta}(\bar{\eta}_\gamma)^m(\bar{\eta}_\delta)^n
\prod^{M_1}_{j\neq i}(\eta_\gamma-z_j)(\eta_\delta-z_j)\\[3mm]
=&\,(M_1-m)\sum^N_{\gamma\delta}(\bar{\eta}_\gamma)^m(\bar{\eta}_\delta)^n
\prod^{M_1}_{i=1}(\eta_\gamma-z_i)(\eta_\delta-z_i)\\
&\,-\sum_{\ell=1}^{\ell_\mathrm{m}}(m-n+2\ell)
\sum^N_{\gamma\delta}(\bar{\eta}_\gamma)^{m+\ell}
(\bar{\eta}_\delta)^{n-\ell}
\prod^{M_1}_{i=1}(\eta_\gamma-z_i)(\eta_\delta-z_i).
\end{split}
\label{eq:twocoltheorem}
\end{equation}
\end{theo}
A proof of Theorem~\ref{theo:twocolorons} is given in
App.~\ref{app:twocolorontheorem}.

After collecting all terms, we obtain for the action of the
Hamiltonian on the momentum eigenstates (\ref{eq:2momcol}):
\begin{equation}
\begin{split}
H_\text{SU(3)}\ket{\Psi_{mn}}=&
-\frac{\pi^2}{18}\left(N-\frac{17}{N}+\frac{52}{N^2}\right)\ket{\Psi_{mn}}\\
&+\frac{3\pi^2}{N^2}\left[(M_1-m)m+(M_1-n)n-\frac{2}{3}(m-n)\right]
\ket{\Psi_{mn}}\\
&-\frac{4\pi^2}{N^2}\sum_{\ell=1}^{\ell_\mathrm{m}}(m-n+2\ell)
\ket{\Psi_{m+\ell,n-\ell}}.
\end{split}
\label{eq:actiononpmn}
\end{equation}
This Sutherland-type equation~\cite{Sutherland71pra,Sutherland72}
shows first of all that when we act with the Hamiltonian on the
non-orthogonal basis state (\ref{eq:2momcol}) with $m=m_0$, $n=n_0$,
$m_0\ge n_0$, we only obtain terms (\ref{eq:2momcol}) with $m\ge m_0$
and $n\le n_0$ with fixed $m+n=m_0+n_0$, but no terms with $m<m_0$ or
$n>n_0$.  Consequently, states with $m=M_1$ or $n=0$ (or both) are
exact eigenstates.  Furthermore, we can obtain all the other exact
eigenstates at each fixed $m+n$ (or fixed total momentum) by
subsequently constructing an orthogonal basis of states starting from
these exact eigenstates.  (We have shown
previously~\cite{GreiterSchurichtsiprb} that this method can be
used to obtain all the spin polarized two spinon eigenstates of the
SU(2) HSM.)

With all the numerical constants in place, however,
(\ref{eq:actiononpmn}) provides much more information than the
direction of the scattering.  We can solve for the energy eigenstates
directly through the Ansatz
\begin{equation}
\Phi_{mn}=\sum_{\ell=0}^{\ell_\mathrm{m}}a^{mn}_{\ell}\Psi_{m+\ell,n-\ell}.
\label{eq:twocolenergyes}
\end{equation}
Requiring
\begin{equation}
H_\text{SU(3)}\ket{\Phi_{mn}}=E_{mn}\ket{\Phi_{mn}}
\label{eq:twocolschroed}
\end{equation}
then yields the recursion relation
\begin{equation}
a^{mn}_{\ell}=\frac{(\ell-\frac{5}{3})(m-n+\ell-1)(m-n+2\ell)}
{\ell(m-n+\ell+\frac{2}{3})(m-n+2\ell-2)}a^{mn}_{\ell-1},\quad a^{mn}_0=1,
\end{equation}
($a^{mm}_1=-4/5$) for the coefficients $a^{mn}_{\ell}$ as well as 
\begin{equation}
E_{mn}=-\frac{\pi^2}{18}\left(N-\frac{17}{N}+\frac{52}{N^2}\right)
+\frac{3\pi^2}{N^2}\left[(M_1-m)m+(M_1-n)n-\frac{2}{3}(m-n)\right]\!.
\label{eq:2colenergy}
\end{equation}
for the two-coloron energies.

As we introduce single-coloron momenta according to 
\begin{equation}
p_m=\frac{4}{3}\pi -\frac{2\pi}{N}\biggl(m+1\biggr),\quad
p_n=\frac{4}{3}\pi -\frac{2\pi}{N}\left(n+\frac{1}{3}\right)\!,
\label{eq:singlecoloronmomenta}
\end{equation}
the energy (\ref{eq:2colenergy}) simplifies to 
\begin{equation}
E_{mn}\,=\,E_0+\frac{4}{9}\frac{\pi^2}{N^2}\,
+\,\epsilon({p_m})\,+\,\epsilon({p_n})
\label{eq:2colenergywithp}
\end{equation}
with $E_0$ given by (\ref{eq:su3-gsenergyM=N/3}) and the
single-coloron dispersion relation $\epsilon(p)$ given by
(\ref{eq:singlecolorondispersion}).  The interpretation of
(\ref{eq:singlecoloronmomenta}) and (\ref{eq:2colenergywithp}) is as
follows. First, the coloron excitations in the SU(3) HSM are (like the
spinon excitations in the SU(2) model~\cite{GreiterSchurichtsiprb})
free, and the total energy is simply given by the sum of the kinetic
energies of the individual colorons.  This conclusion is consistent
with the asymptotic Bethe Ansatz calculation by
Essler~\cite{Essler95}, who has shown that the coloron-coloron
scattering matrix is trivial, \ie a momentum independent phase.
Second, the difference in the individual coloron momenta
(\ref{eq:singlecoloronmomenta}) is quantized as
\begin{equation}
  \label{eq:coloronmomentaspacing}
  p_n-p_m=\frac{2\pi}{N}\left(\frac{2}{3}+\text{integer}\right),
\end{equation}
\ie the minimal momentum spacing between color-polarized colorons is
${2}/{3}$ of the momentum spacing for fermions on a ring, $2\pi/N$.
We interpret this result as a manifestation of the fractional
statistics of the colorons.  This result appears to indicate that
the spacing for particles with exclusion statistics $g$ is given
by 
\begin{equation}
  \label{eq:momentaspacing}
  p_n-p_m=\frac{2\pi}{N}\left(g+\text{integer}\right).
\end{equation}
This identification is obviously consistent with the familiar cases of
bosons and fermions, and may even constitute the most direct
manifestation of Haldane's~\cite{Haldane91prl2} fractional exclusion
principle {\it per se}.

The fractional momenta (\ref{eq:singlecoloronmomenta}) may also be
interpreted in terms of an effective change in the periodic boundary
conditions (PBCs).  The allowed values for the total momenta of
the two-coloron states are those for conventional PBCs,
\begin{equation}
  \label{eq:colorontotalmomenta}
  p_n+p_m=\frac{2\pi}{N}\cdot\text{integer}
\end{equation}
where we have used the fact that $N+2$ in (\ref{eq:singlecoloronmomenta}) 
is divisible by three.  This is also required, as the wave
function must come back to itself under a full translation around the
ring.  The allowed values for the difference in the momenta
(\ref{eq:coloronmomentaspacing}), by contrast, are those of a ring
threaded by a magnetic flux $2\pi\!\cdot\!2/3$.  In a certain sense,
we may interpret this flux as a phase acquired by the wave function
as one coloron goes ``through'' the other.  There are, however,
several subtleties associated with this
interpretation~\cite{manuscriptinpreparationGST}.

\section{SU(3) spin currents}

In this section, we calculate the SU(3) spin currents, \ie the
eigenvalues of the diagonal rapidity operators $\Lambda^3$ and
$\Lambda^8$.  To this end, we first rewrite them in terms of colorflip
operators:
\begin{eqnarray}
\Lambda^3&=&-\frac{1}{2}\sum_{\alpha\neq\beta}^N
\frac{\eta_\alpha+\eta_\beta}{\eta_\alpha-\eta_\beta}
\left(e_\alpha^{\b\r}e_\beta^{\r\b}+
\frac{1}{2}e_\alpha^{\b\g}e_\beta^{\g\b}-
\frac{1}{2}e_\alpha^{\r\g}e_\beta^{\g\r}\right)\!,\\
\Lambda^8&=&-\frac{\sqrt{3}}{4}\sum_{\alpha\neq\beta}^N
\frac{\eta_\alpha+\eta_\beta}{\eta_\alpha-\eta_\beta}
\left(e_\alpha^{\b\g}e_\beta^{\g\b}+
e_\alpha^{\r\g}e_\beta^{\g\r}\right)\!.
\end{eqnarray}
The eigenvalue of $\Lambda^8$ is easier to evaluate. 
Using the Taylor expansion technique we obtain
\begin{eqnarray}
\left[\Lambda^8\Psi\right]\!\![z_i;w_k]&=&
\frac{\sqrt{3}}{4}\sum_{m=0}^{N-2}\frac{B_m}{m!}
\left[\sum_{i=1}^{M_1}z_i^{m+1}\frac{\partial^m}{\partial z_i^m}
\frac{\Psi}{z_i}+\sum_{k=1}^{M_2}w_k^{m+1}\frac{\partial^m}{\partial w_k^m}
\frac{\Psi}{w_k}\right]\nonumber\\[2mm]
&=&\frac{\sqrt{3}}{4}(N-2)(M_1+M_2)\Psi-
\frac{\sqrt{3}}{2}\left[\sum_{i=1}^{M_1}z_i^2\frac{\partial}{\partial z_i}
\frac{\Psi}{z_i}+\sum_{k=1}^{M_2}w_k^2\frac{\partial}{\partial w_k}
\frac{\Psi}{w_k}\right]\!,\label{eq:spincurrent}
\end{eqnarray}
where the wave function has to satisfy $\text{deg}_{z_i}\Psi\le N-1$
and $\text{deg}_{w_k}\Psi\le N-1$, and we have defined
$B_m\equiv\sum_{\alpha=1}^{N-1}
\eta_\alpha(\eta_\alpha+1)(\eta_\alpha-1)^{m-1}$.  Evaluation of the
latter yields $B_0=N-2$, $B_1=-2$, and $B_m=0$ for $2\le m\le N-2$
(see App.~\ref{app:formulas}).

\begin{figure}[t]
\includegraphics[scale=0.20]{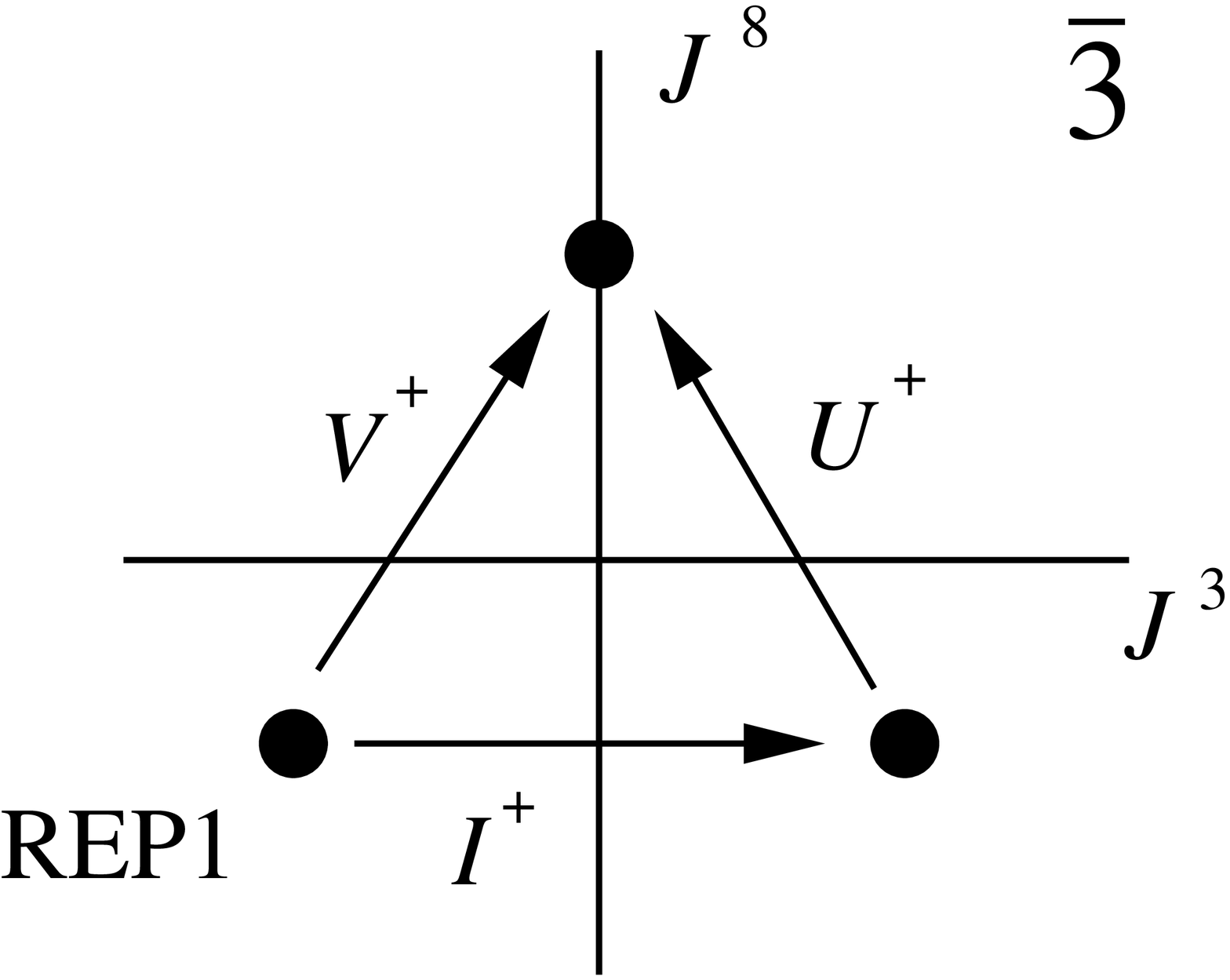}\hspace{10mm}
\includegraphics[scale=0.20]{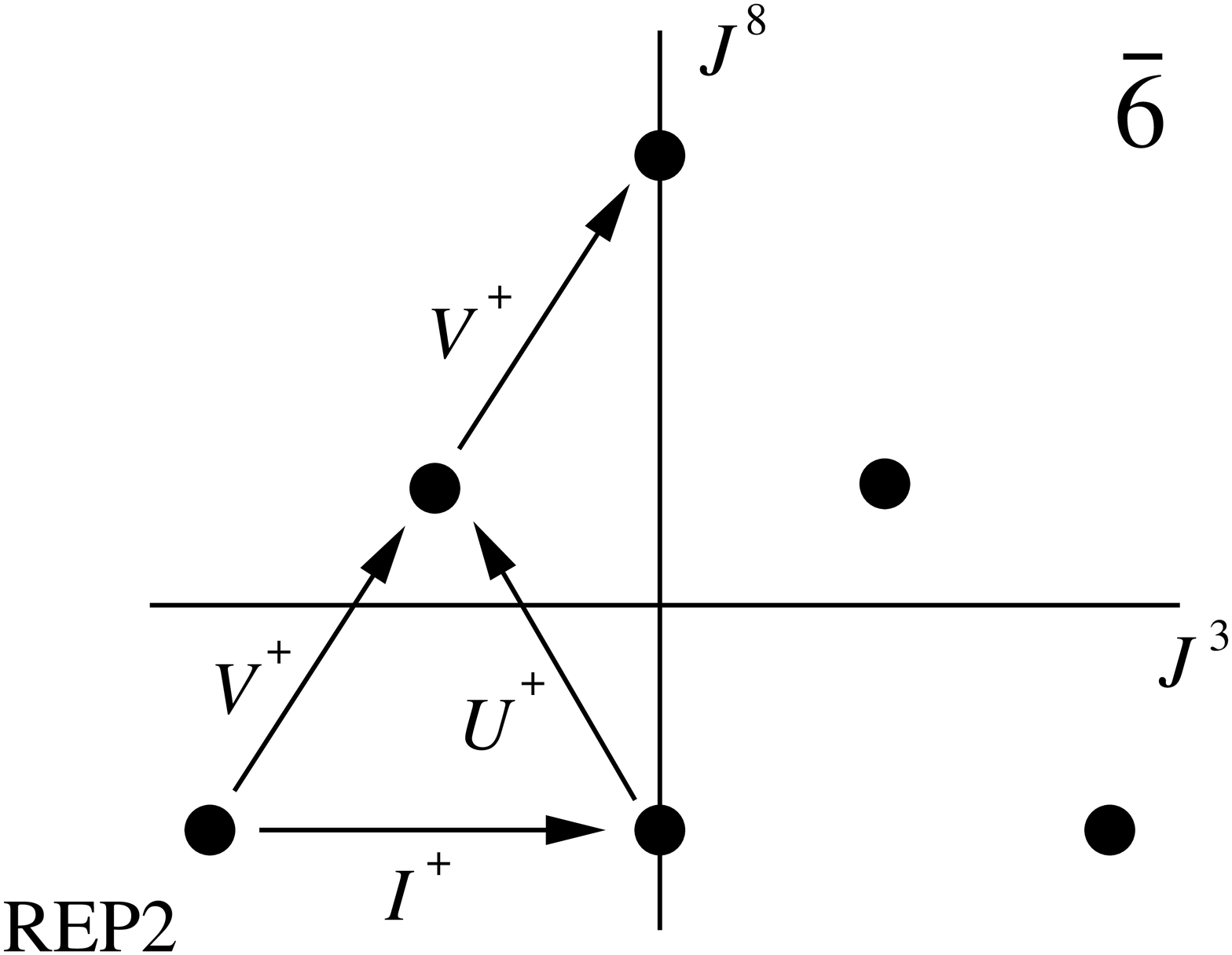}
\caption{Weight diagrams of the SU(3) representations 
  $\bar{3}$ and $\bar{6}$.}
\label{fig:lwsreps}
\end{figure}

With (\ref{eq:spincurrent}) we find for the ground state, the
one-coloron states (\ref{eq:wfmomentumcoloron}), and the two-coloron
states (\ref{eq:twocolenergyes})
\begin{eqnarray}
\Lambda^8\ket{\Psi_0}&=&0,\label{eq:sc1}\\[2mm]
\Lambda^8\ket{\Psi_n}&=&-\frac{\sqrt{3}}{4}\left(\frac{N-2}{3}-2n\right)
\ket{\Psi_n},\label{eq:sc2}\\[2mm]
\Lambda^8\ket{\Phi_{mn}}&=&
-\frac{\sqrt{3}}{4}\left(\frac{2N-8}{3}-2m-2n\right)
\ket{\Phi_{mn}}\label{eq:sc3}.
\end{eqnarray}
The eigenvalues of $\Lambda^3$ can be obtained either by explicit
calculation or be determined as follows.  The fundamental
representations of the Yangian $Y(\mathrm{sl}_3)$ can be
constructed~\cite{ChariPressley98} from the representations of
$\mathrm{sl}_3$ (\ie SU(3)) by the pull-back under the so-called
evaluation homomorphism from $Y(\mathrm{sl}_3)$ into the universal
enveloping algebra $U(\mathrm{sl}_3)$ of $\mathrm{sl}_3$.  Explicitly,
the ground state transforms according to the pull-back of the SU(3)
singlet representation under $Y(\mathrm{sl}_3)$ transformations, the
one-coloron states transform according to the pull-back of the
representation $\bar{3}$, and the two-coloron states investigated in
Sec.~\ref{sec:twocolorons} according to the pull-back of the
representation $\bar{6}$ (as the two colorons are coupled
symmetrically).  For the fundamental representations of
$Y(\mathrm{sl}_3)$ constructed in this way the ratio of the
eigenvalues of $\Lambda^3$ and $\Lambda^8$ is equal to the ratio of
the eigenvalues of $J^3$ and $J^8$. Hence, as the states $\ket{\Psi_n}$ and
$\ket{\Phi_{mn}}$ are SU(3) lowest weight states (\ie they are annihilated
by $I^-=J^1-iJ^2$, $V^-=J^4-iJ^5$, and $U^-=J^6-iJ^7$; see
Fig.~\ref{fig:lwsreps}), the eigenvalues of $\Lambda^3$ are simply
$\sqrt{3}$ times the eigenvalues of $\Lambda^8$ given in
(\ref{eq:sc1})--(\ref{eq:sc3}).  
\psfrag{REP1}{$\ket{\Psi_n}$}
\psfrag{REP2}{$\ket{\Phi_{mn}}$}

\section{Asymptotic Bethe Ansatz}

Let us now briefly compare our results with conclusions drawn from the
asymptotic Bethe Ansatz (ABA).  We wish to stress that the ABA is not an exact
method. In the derivation of the ABA equations one assumes the existence of an
asymptotic region in which the particles (or colorflips) are
non-interacting~\cite{Kawakami92prb1}; an assumption which is not valid
rigorously. In particular, the ABA equations have no direct connection to the
exact eigenstates of the Hamiltonian~\cite{Essler95} as it is the case for the
standard Bethe Ansatz. It turns out, nevertheless, that the ABA reproduces the
spectrum of the model, hence there is an indirect relation between solutions
of the ABA equations and energy eigenstates. Let us now set up the formalism.
The eigenstates of the SU(3) HSM are specified by two sets of pseudomomenta
$k_i^1$ and $k_i^2$ satisfying the ABA
equations~\cite{Kawakami92prb1,HaHaldane93}
\begin{equation}
\begin{split}
&Nk_i^1=2\pi I^1_i-\pi\sum_{j=1}^{M^2}\text{sgn}(k_i^1-k_j^2)
+\pi\sum_{j=1}^{M^1}\text{sgn}(k_i^1-k_j^1),\\
&\pi\sum_{j=1}^{M^1}\text{sgn}(k_i^2-k_j^1)=2\pi I^2_i+
\pi\sum_{j=1}^{M^2}\text{sgn}(k_i^2-k_j^2),
\end{split}
\label{eq:ABAequations}
\end{equation}
with two sets of mutually distinct integer (or half-integer) quantum numbers
in the range
\begin{equation}
|I^1_i|\le \frac{1}{2}(N+M^2-M^1-1),\quad
|I^2_i|\le \frac{1}{2}(M^1-M^2-1).
\end{equation}
In (\ref{eq:ABAequations}) we have restricted ourselves to 1-strings,
\ie the $k_i^{1,2}$'s are real numbers, which means that we can only
describe color-polarized states (for the general case
see~\cite{HaHaldane93,Essler95}).  The energy and momentum depend only
on the first set of pseudomomenta,
\begin{eqnarray}
  E&=&\frac{1}{4}\sum_{i=1}^{M^1}\Bigl((k_i^1)^2-\pi^2\Bigr)+
  \frac{\pi^2}{N^2}\frac{N(N^2-1)}{18},\\
  p&=&\sum_{i=1}^{M^1}(k_i^1+\pi)\ \ \text{mod}\;2\pi,
\end{eqnarray}
where $k_i^1\in [-\pi,\pi]$.  

The ground state is obtained by arranging the $I^{1,2}_i$'s densely
around zero, \ie by choosing $M^1=2M$ and $M^2=M$ for a chain of
length $N=3M$.  Excitations are created by introducing holes into
these sets of quantum numbers~\cite{Andrei92}, \eg for $N=3M-1$ we
obtain a single hole $I_\mathrm{h}^2$ in the rank-2 quantum numbers by
choosing $M^1=2M-1$ and $M^2=M-1$.  In all practical applications we
have to take the thermodynamic limit, such that the ABA equations
(\ref{eq:ABAequations}) turn into integral equations.  We introduce
pseudomomenta densities
\begin{equation}
\sigma^1(k_i^1)=\frac{1}{k^{1}_{i+1}-k^{1}_i},\quad
\sigma^2(k_i^2)=\frac{1}{k^{2}_{i+1}-k^{2}_i},
\end{equation} 
as well as bare hole densities
$\sigma^{1,2}_\text{h}=\sum_j\delta(k-\lambda^{1,2}_j)$ which describe
the excitations. Then the ABA equations (\ref{eq:ABAequations}) turn
into integral equations with $\delta$-functions as integral kernels
\begin{equation}
\begin{split}
  \sigma^1(k)+\sigma^1_\text{h}(k)&=
  \frac{N}{2\pi}-\sigma^1(k)+\sigma^2(k),\\
  \sigma^2(k)+\sigma^2_\text{h}(k)&=\sigma^1(k)-\sigma^2(k).
\label{eq:sigmaeq}
\end{split}
\end{equation}
The latter are a special and distinguishing feature of the HSM.  For
the ground state we have $\sigma^{1,2}_\text{h}=0$ and
(\ref{eq:sigmaeq}) is solved by $\sigma_0^1=N/3\pi$ and
$\sigma_0^2=N/6\pi$~\cite{Essler95}. This shows the simple nature of
the ground state, which is explicitly given by the Gutzwiller state
(\ref{eq:su3-nnhgroundstate}). It is, however, not possible to
derive the Gutzwiller wave function from the ground state densities
$\sigma_0^{1,2}$ as there is no direct connection between solutions of
the ABA equations and the exact eigenstates of the Hamiltonian.

One-coloron states correspond to single holes in the rank-2
pseudomomenta, \ie $\sigma^1_\text{h}=0$ and
$\sigma^2_\text{h}=\delta(k-\lambda^2)$, which implies
\begin{equation}
\sigma^1(k)=\sigma_0^1-\frac{1}{3}\sigma^2_\text{h},\quad
\sigma^2(k)=\sigma_0^2-\frac{2}{3}\sigma^2_\text{h}.
\label{eq:sigmacoloron}
\end{equation}
There is no ``hole-dressing'' in the surrounding pseudomomenta, or
equivalently, the pseudomomenta densities are only affected through trivial
shifts as we introduce holes.  This observation reflects once again the fact
that colorons in the SU(3) HSM (as well as spinons in the HSM) are
free~\cite{GreiterSchurichtsiprb}. The densities (\ref{eq:sigmacoloron}) yield
the correct coloron momenta (\ref{eq:coloronmomentum}) and energies
(\ref{eq:colorondisperion}) in the thermodynamic limit.  Holes in the rank-2
pseudomomenta represent excitations which transform under
$\bar{3}$~\cite{Essler95}, hence we find again that colorons possess
complementary colors~\cite{SchurichtGreiter05colepl}. We wish to stress,
however, that this result cannot be derived using the ABA exclusively. First,
it cannot be answered whether rank-1 or rank-2 holes represent elementary
excitation, and second, the ABA is not an exact method in contrast to
calculations based on explicit wave functions.  From (\ref{eq:sigmacoloron})
we can further read off that when adding one coloron, the number of available
orbitals described by $\sigma^2(k)$ is reduced by $2/3$. It is, however, not
completely clear to us whether this can be interpreted directly as a
derivation of the exclusion statistics of polarized colorons.

\section{Generalization to SU($\bs{n}$) }

The results derived here generalize directly to the SU($n$) HSM.
Consider a chain with one particle per lattice site carrying an
internal SU($n$) quantum number which transforms according to the
fundamental representation n of SU($n$).  The Hamiltonian is given by
\begin{equation}
H_{\text{SU($n$)}}=\;\frac{2\pi^2}{N^2}
\sum^N_{\alpha<\beta}\sum^n_{\sigma\tau}
\frac{e_\alpha^{\sigma\tau}e_\beta^{\tau\sigma}}
{\vert\eta_\alpha -\eta_\beta\vert^2},
\label{eq:su3-sunhamiltonian}
\end{equation}
where the operator $e_\alpha^{\sigma\tau}$ annihilates a particle of
flavor $\tau$ at site $\eta_\alpha$ and creates a particle of flavor
$\sigma$ at the same site.  (Once again, $\tau$ and $\sigma$ may be
equal in (\ref{eq:su3-sunhamiltonian}).  In comparison to
(\ref{eq:su3hamiltonian}), we have omitted a constant.)  The model
again possesses a Yangian
symmetry~\cite{HaldaneHaTalstraBernardPasquier92} and hence
is integrable.

If we use a polarized state of particles of flavor $n$ as reference
state and label the coordinates of the particles of flavor $\sigma,
1\le\sigma\le n-1$, by $z_i^\sigma, 1\le i \le M_\sigma$, the wave
functions~\cite{Kawakami92prb2}
\begin{equation}
\Psi_0[z_i^\sigma]=\prod_{\sigma=1}^{n-1}
\prod^{M_\sigma}_{i<j}(z_i^\sigma-z_j^\sigma)^2
\prod^{n-1}_{\sigma<\tau}
\prod_{i=1}^{M_\sigma}\prod_{j=1}^{M_\tau}(z_i^\sigma-z_j^\tau)
\prod_{\sigma=1}^{n-1}\prod_{i=1}^{M_\sigma}z_i^\sigma
\label{eq:su3-definitionsunpsi0}
\end{equation}
constitute exact eigenstates~\cite{HaHaldane92} of the Hamiltonian
(\ref{eq:su3-sunhamiltonian}).  For $N=nM$, $M_\sigma=M$,
(\ref{eq:su3-definitionsunpsi0}) is the ground state of
(\ref{eq:su3-sunhamiltonian}) with energy~\cite{Kawakami92prb2}
\begin{equation}
E_0^n=-\frac{\pi^2}{12}\left(\frac{n-2}{n}N+\frac{2n-1}{N}\right)\!.
\end{equation}
The total momentum of this ground state is $p=(n-1)\pi
M\!\!\mod{2\pi}$.

\begin{figure}[t]
 \setlength{\unitlength}{10pt}
 \begin{picture}(16,9)(-1,0)
 \put(-3,8){a) $n$ even}
 \put(0,1){\line(1,0){13.5}}
 \put(0,1){\line(0,1){5}}
 \multiput(12,1.3)(0,0.6){7}{\rule{0.3pt}{3pt}}
 \qbezier[30](0,4.5)(0.8,4.5)(1.5,1)
 \qbezier[2000](4.5,1)(6,8)(7.5,1)
 \qbezier[30](10.5,1)(11.2,4.5)(12,4.5)
 \put(1.5,1){\makebox(0,0)[t]{\rule{0.3pt}{4pt}}}
 \put(4.5,1){\makebox(0,0)[t]{\rule{0.3pt}{4pt}}}
 \put(7.5,1){\makebox(0,0)[t]{\rule{0.3pt}{4pt}}}
 \put(10.5,1){\makebox(0,0)[t]{\rule{0.3pt}{4pt}}}
 \put(12,1){\makebox(0,0)[t]{\rule{0.3pt}{4pt}}}
 \put(0,0){\makebox(0,0){\small $0$}}
 \put(12,0){\makebox(0,0){\small $2\pi$}}
 \put(13.7,0.4){\makebox(0,0){\small $p$}}
 \put(-1.3,6){\makebox(0,0){\small $\epsilon(p)$}}
 \put(0.75,1){\makebox(0,0){\rule{15pt}{3pt}}}
 \put(6,1){\makebox(0,0){\rule{30pt}{3pt}}}
 \put(11.25,1){\makebox(0,0){\rule{15pt}{3pt}}}
 \end{picture}
\hspace{20mm}
 \begin{picture}(16,9)(-1,0)
 \put(-3,8){b) $n$ odd}
 \put(0,1){\line(1,0){13.5}}
 \put(0,1){\line(0,1){5}}
 \qbezier[2000](4.5,1)(6,8)(7.5,1)
 \put(4.5,1){\makebox(0,0)[t]{\rule{0.3pt}{4pt}}}
 \put(7.5,1){\makebox(0,0)[t]{\rule{0.3pt}{4pt}}}
 \put(12,1){\makebox(0,0)[t]{\rule{0.3pt}{4pt}}}
 \put(0,0){\makebox(0,0){\small $0$}}
 \put(12,0){\makebox(0,0){\small $2\pi$}}
 \put(13.7,0.4){\makebox(0,0){\small $p$}}
 \put(-1.3,6){\makebox(0,0){\small $\epsilon(p)$}}
 \put(6,1){\makebox(0,0){\rule{30pt}{3pt}}}
 \end{picture}
 \caption{SU($n$) spinon dispersion relations. a) $n$ even. The allowed 
   momenta fill the interval $[\pi-\frac{\pi}{n},\pi+\frac{\pi}{n}]$ for
   $M$ even and $[-\frac{\pi}{n},\frac{\pi}{n}]$ for $M$ odd. b) $n$ odd. 
   The allowed momenta fill the interval 
   $[\pi-\frac{\pi}{n},\pi+\frac{\pi}{n}]$.}
 \label{fig:disn}
 \end{figure}
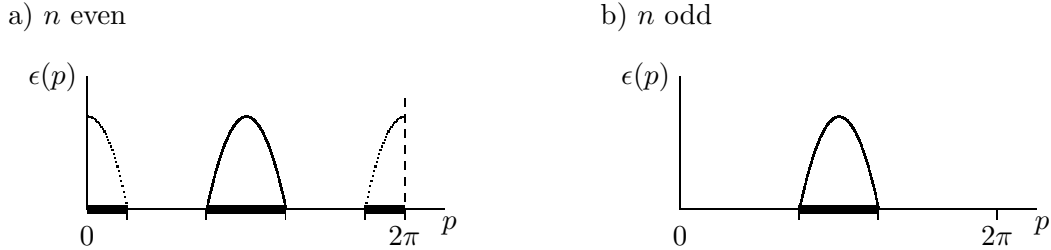

Localized SU($n$) spinons are given by
\begin{equation}
  \Psi_\gamma[z_i^\sigma]=\prod_{i=1}^{M_1}(\eta_\gamma-z_i^1)\;
  \Psi_0[z_i^\sigma]
  \label{eq:su3-localspinon}
\end{equation}
for $N=nM-1$, $M_1=M-1$, and $M_2=\ldots=M_{n-1}=M$. As hole-like
excitations they transform according to the representation
$\bar{\text{n}}$.  Energy eigenstates are obtained by Fourier
transformation of (\ref{eq:su3-localspinon}),
\begin{equation}
  \Psi_m[z_i^\sigma]=\frac{1}{N}\sum_{\gamma=1}^N
  (\bar{\eta}_\gamma)^m\Psi_{\gamma}[z_i^\sigma],\quad
  0\le m \le M_1.
  \label{eq:momsunspinon}
\end{equation}
Their momenta are given by
\begin{equation}
  p=\frac{n-1}{n}\pi N-\frac{2\pi}{N}\left(m+\frac{n-1}{2n}\right)
  \!\!\mod{2\pi},
\end{equation}
which fill the interval $[-\frac{\pi}{n},\frac{\pi}{n}]$ for $n$ even
and $M$ odd, or the interval $[-\frac{\pi}{n},\frac{\pi}{n}]$
otherwise (either $n$ odd or $M$ even or both).  The energy
eigenvalues of the one SU($n$) spinon states (\ref{eq:momsunspinon})
are given by
\begin{equation}
  E^n(p)=E_0^n+\frac{n^2-1}{12n}\frac{\pi^2}{N^2}\,+\,\epsilon^n(p)
\end{equation}
with (see Fig.~\ref{fig:disn})
\begin{equation}
  \label{eq:epsilonpn}
  \epsilon^n(p)=\left\{ 
    \begin{array}{l@{\hspace{15pt}}l}
      \displaystyle
      \frac{n}{4}\left(\frac{\pi^2}{n^2}-p^2\right)\!, 
      & \text{if $n$ even and $M$ odd},\\[15pt]
      \displaystyle
      \frac{n}{4}\left(\frac{\pi^2}{n^2}-(p-\pi)^2\right)\!, 
      & \text{otherwise}.
    \end{array}\right.
\end{equation}

The statistical parameter for polarized SU($n$) spinons is
$g=(n-1)/n$.  In the framework of the ABA, the spinons are represented
by holes in rank-$(n-1)$ pseudomomenta.

Finally, let us add some remarks on the large-$n$ limit of the SU($n$) HSM.
First, there are no simplifications in the calculations which yield the
results presented in this section. When $n$ grows, we obtain solely terms
similar to the ones treated in the appendix. Second, the interval of allowed
spinon momenta shrinks to zero. This does not mean that the number of spinon
orbitals vanishes, as in the limit $n\rightarrow\infty$ the number of lattice
sites $N=nM-1$ has to grow and hence the momentum spacing tends to zero as
well.  Third, we find $g\rightarrow 1$ as $n\rightarrow\infty$, meaning that
the exclusion statistics between polarized spinons becomes fermionic in this
limit. This does not imply, however, that spinons in the SU($n$) HSM behave
like free fermions in the large-$n$ limit. Their momenta fill only a narrow
interval and the individual spinon momenta are not simply given by integer
multiples of $2\pi/N$~\cite{manuscriptinpreparationGS}.

\section{Conclusion}

In conclusion, we have shown by explicit calculation that the elementary
excitations of SU($n$), but in particular SU(3), HSM transform under the
representation conjugate to the representation of the SU($n$) spins on the
chain.  We have also shown that they obey fractional statistics, and only
exist in one n$^\text{th}$ of the Brillouin zone.  We strongly believe these
features to be generic features of ``antiferromagnetic'' SU($n$) spin chains.
We have further endeavored to construct the polarized two-coloron eigenstates
of the HSM explicitly, and have thereby shown that they are free and that the
spacings in the difference of the individual coloron momenta $p_n$ and $p_m$
with $p_n\ge p_m$ are given by
$p_n-p_m=\frac{2\pi}{N}\left(\frac{2}{3}+\text{integer}\right)$.  We have
interpreted this peculiar shift as a manifestation of the fractional
statistics.  This indicates that fractional statistics in one dimension
manifests itself not only through a fractional exclusion
principle~\cite{Haldane91prl2}, but that there are Berry's phases associated
with anyons~\cite{Wilczek1990} in one dimensions as
well~\cite{manuscriptinpreparationGST}. Furthermore, we look forward to an
experimental realization of SU(3) spin chains in systems of ultracold
atoms~\cite{AbrahamMcAlexanderGertonHuletCoteDalgarno97,RegalJin03}, and hence
an observation of the complementary colors of the excitations.

\section*{ACKNOWLEDGMENTS}
MG would like to thank the organizers of the 2003 Amsterdam Summer
Workshop on Flux, Charge, Topology and Statistics, where this work was
partially inspired.  DS was supported by the German Research
Foundation (DFG) through GK 284.

\appendix
\section{Gell-Mann matrices}\label{app:conventions}

The Gell-Mann matrices are given by~\cite{Georgi82}
\begin{displaymath}
\begin{split}
&\lambda^1=\left(\begin{array}{ccc}0&1&0\\1&0&0\\0&0&0\end{array}
\right)\!\!,\quad
\lambda^2 =\left(\begin{array}{ccc}0&-i&0\\i&0&0\\0&0&0\end{array}
\right)\!\!,\quad
\lambda^3 =\left(\begin{array}{ccc}1&0&0\\0&-1&0\\0&0&0\end{array}
\right)\!\!,\\[3mm]
&\lambda^4=\left(\begin{array}{ccc}0&0&1\\0&0&0\\1&0&0\end{array}
\right)\!\!,\quad
\lambda^5 =\left(\begin{array}{ccc}0&0&-i\\0&0&0\\i&0&0\end{array}
\right)\!\!,\quad
\lambda^6 =\left(\begin{array}{ccc}0&0&0\\0&0&1\\0&1&0\end{array}
\right)\!\!,\\[3mm]
&\lambda^7=\left(\begin{array}{ccc}0&0&0\\0&0&-i\\0&i&0\end{array}
\right)\!\!,\quad
\lambda^8 =\frac{1}{\sqrt{3}}
\left(\begin{array}{ccc}1&0&0\\0&1&0\\0&0&-2\end{array}\right)\!\!.
\end{split}
\end{displaymath}
They are normalized as
$\mathrm{tr}\left(\lambda^a\lambda^b\right)=2\delta_{ab}$ and satisfy the
commutation relations $\comm{\lambda^a}{\lambda^b}=2f^{abc}\lambda^c.$ The
structure constants $f^{abc}$ are totally antisymmetric and obey Jacobi's
identity
\begin{displaymath}
f^{abc}f^{cde}+f^{bdc}f^{cae}+f^{dac}f^{cbe}=0.
\end{displaymath} 
Explicitly, the non-vanishing structure constants are given by $f^{123}=i$,
$f^{147}=f^{246}=f^{257}=f^{345}=-f^{156}=-f^{367}=i/2$,
$f^{458}=f^{678}=i\sqrt{3}/2$, and 45 others obtained by permutations of 
the indices.
The SU(3) spin operators are expressed in terms of the colorflip operators by
\begin{displaymath}
\bs{J}_\alpha\!\cdot\!\bs{J}_\beta\equiv
\sum_{a=1}^8 J^a_\alpha J^a_\beta=
\frac{1}{2}\sum_{\sigma\tau}^3\,e_\alpha^{\sigma\tau}\,e_\beta^{\tau\sigma} 
-\frac{1}{6}.
\end{displaymath}

\section{Equivalence of (\ref{eq:su3-nnhgroundstate}) 
and (\ref{eq:su3-definitionpsi0})}\label{app:gwstate}

Let us begin by exploring a general consequence of particle-hole
symmetry. Consider a chain of $N$ sites.  The wave function of $L$
fermions with momenta in the interval $I_1$ is given by
\begin{equation}
  \Phi_1[u_1,\ldots,u_L]=\bra{0}c_{u_1}\cdots c_{u_L}
  \prod_{q\in I_1}c_q^\dagger\ket{0}\!.
  \label{eq:defphi}
\end{equation}
This wave function can be expressed in terms of $N-L$ fermions whose
momenta fill the interval $I_2=[-\pi,\pi]\backslash I_1$, 
\begin{eqnarray}
  \Phi_1[u_1,\ldots,u_L]&=&\biggl(\bra{0}\prod_{q\in I_1}c_q\; 
  c_{u_L}^\dagger\cdots c_{u_1}^\dagger\ket{0}\biggr)^{\!\!*}\nonumber\\
  &=&\text{sign}\cdot\biggl(\bra{0}\prod_{q\in I_1}c_q\; 
  c_{v_1}\cdots c_{v_{N-L}}\prod_{q\in I_2}c_q^\dagger
  \prod_{q\in I_1}c_q^\dagger\ket{0}\biggr)^{\!\!*}\nonumber\\[2mm]
  &=&\text{sign}\cdot\Phi_2^*[v_1,\ldots,v_{N-L}]\label{eq:etazsymmetry},
\end{eqnarray}
where the coordinates $v_1,\ldots,v_{N-L}$ are those coordinates
on the chain which are not contained in the set $u_1,\ldots,u_{L}$,
\begin{equation}
  \Phi_2[v_1,\ldots,v_{N-L}]=\bra{0}c_{v_1}\cdots c_{v_{N-L}}
  \prod_{q\in I_2}c_q^\dagger\ket{0}\!,
\label{eq:defphi2}
\end{equation}
and the overall sign depends on the number and positions of the
fermions.

(\ref{eq:etazsymmetry}) enables us to show that the ground
state wave function is given by (\ref{eq:su3-definitionpsi0}) with
$M_1=M_2=M$. For $N=3M$, the wave function of the Slater determinant
state $\ket{\Psi_{\text{SD}}}$ is (see Fig.~\ref{fig:sd}a)
\begin{equation}
\Psi_\mathrm{SD}[z_i;w_k;u_m]=
\Phi_1[z_i]\Phi_1[w_k]\Phi_1[u_m],
\label{eq:sdwf}
\end{equation}
where the interval $I_1$ is chosen to be $[-\pi/3,\pi/3]$ and hence
\begin{equation}
\Phi_1[z_i]=\prod_{i=1}^Mz_i^{-\frac{M-1}{2}}\prod_{i<j}^M(z_i-z_j).
\label{eq:sdwf1}
\end{equation}
The prefactor in (\ref{eq:sdwf1}) shifts all momenta from the interval
$[0,2\pi/3]$ to $I_1$. The ground state of the SU(3) HSM is obtained
by Gutzwiller projection,
$\ket{\Psi_0}=P_\text{G}\ket{\Psi_\text{SD}}$, which enforces that all
coordinates $z_i$, $w_k$, and $u_m$ are distinct.  Hence, we can
express the coordinates $u_m$ in terms of the $z_i$'s and $w_k$'s 
using (\ref{eq:etazsymmetry}),
\begin{displaymath}
\Phi_1[u_m]=\text{sign}\cdot\Phi_2^*[z_i,w_k],
\end{displaymath}
where the wave function of the filled band with momenta in the interval
$I_2=[-\pi,\pi]\backslash I_1$ is (up to a sign) given by
\begin{equation}
\Phi_2[z_i,w_k]=\prod_{i=1}^Mz_i^{\frac{M+1}{2}}
\prod_{k=1}^Mw_k^{\frac{M+1}{2}}\,
\prod_{i<j}^M(z_i-z_j)\prod_{k<l}^M(w_k-w_l)
\prod_{i=1}^M\prod_{k=1}^M(z_i-w_k).
\end{equation}
The ground state wave function is hence given by
\begin{eqnarray}
\Psi_0[z_i,w_k]&=&\Phi_1[z_i]\Phi_1[w_k]\Phi_2^*[z_i,w_k]\nonumber\\[2mm]
&=&\prod_{i=1}^{M}z_i^{-M}\prod_{k=1}^{M}w_k^{-M}
   \prod^{M}_{i<j}|z_i-z_j|^2\prod^{M}_{k<l}|w_k-w_l|^2
   \prod_{i=1}^{M}\prod_{k=1}^{M}(z_i-w_k)^*,\label{eq:apppsi_0}
\end{eqnarray}
where we have used the fact that the coordinates are located on the
unit circle, \eg $z_i^*=1/z_i$.  Using
\begin{displaymath}
|z_i-z_j|^2=-\frac{(z_i-z_j)^2}{z_iz_j}, \quad 
\prod^{M}_{i<j}\frac{1}{z_iz_j}=\prod_{i=1}^{M}\frac{1}{z_i^{M-1}},
\end{displaymath}
as well as
\begin{displaymath}
(z_i-w_k)^*=-\frac{(z_i-w_k)}{z_iw_k}, \quad 
\prod_{i=1}^{M}\prod_{k=1}^{M}\frac{1}{z_iw_k}=
\prod_{i=1}^{M}\frac{1}{z_i^M}\prod_{k=1}^{M}\frac{1}{w_k^M}
\end{displaymath}
we obtain from (\ref{eq:apppsi_0})
\begin{displaymath}
\Psi_0[z_i,w_k]=\prod_{i=1}^{M}z_i^{-3M}\prod_{k=1}^{M}w_k^{-3M}
   \prod^{M}_{i<j}(z_i-z_j)^2\prod^{M}_{k<l}(w_k-w_l)^2
   \prod_{i=1}^{M}\prod_{k=1}^{M}(z_i-w_k)
   \prod_{i=1}^{M}z_i\prod_{k=1}^{M}w_k.
\end{displaymath}
As the coordinates $z_i$ and $w_k$ are located at the lattice sites
$\eta_\alpha=\exp\left(\frac{2\pi i}{N}\alpha\right)$ and $N=3M$, the
prefactor reduces to one and $\Psi_0[z_i,w_k]$ takes the form 
(\ref{eq:su3-definitionpsi0}).

\section{Useful formulas}\label{app:formulas}

Some of the results presented in this appendix can be found
in~\cite{BernevigGiulianoLaughlin01prb}.

\begin{enumerate}
\item 
\begin{equation}
\eta_\alpha^N=1,\quad
\sum_{\alpha=1}^N\eta_\alpha^m=N\,\delta_{0m}, \quad
\prod_{\alpha=1}^N\eta_\alpha=(-1)^{N-1}.
\end{equation}

\item
\begin{equation}
\prod_{\alpha=1}^N(z-\eta_\alpha)=z^N-1.
\end{equation}

\item
\begin{equation}
\sum_{\alpha=1}^N\frac{\eta_{\alpha}}{z-\eta_{\alpha}}=\frac{N}{z^N-1}.
\label{eq:app-hssum5}
\end{equation}
\emph{Proof:} Rewrite (\ref{eq:app-hssum5}) as
\begin{displaymath}
f(z)\equiv \sum_{\alpha=1}^N\eta_{\alpha}
\prod_{\beta\neq\alpha}^N (z-\eta_{\beta})=N,
\end{displaymath}
with $f$ a polynomial of degree $N-1$, $f(0)=N$, and $
f(1/\eta_\gamma)=\prod_{\beta=1}^{N-1}(1-\eta_\beta),\gamma=1,\ldots,
N$.  Since the polynomial $f-f(1)$ has degree $N-1$ and $N$ different
zeros, it vanishes by the fundamental theorem of
algebra~\cite{Lang85}.

\item
The previous statement implies
\begin{equation}
  \sum_{\alpha=1}^N\frac{1}{\eta-\eta_{\alpha}}=
  \frac{N\eta^{N-1}}{\eta^N-1}.
\label{eq:app-hssum6}
\end{equation}

\item
\begin{equation}
  \sum_{\alpha=1}^{N-1}\frac{\eta_\alpha^m}{\eta_\alpha -1}
  =\frac{N+1}{2}-m,
  \quad 1\le m \le N.
  \label{eq:app-hsfouriersum1}
\end{equation}
\begin{figure}[t]
\begin{center}
\includegraphics[scale =0.16]{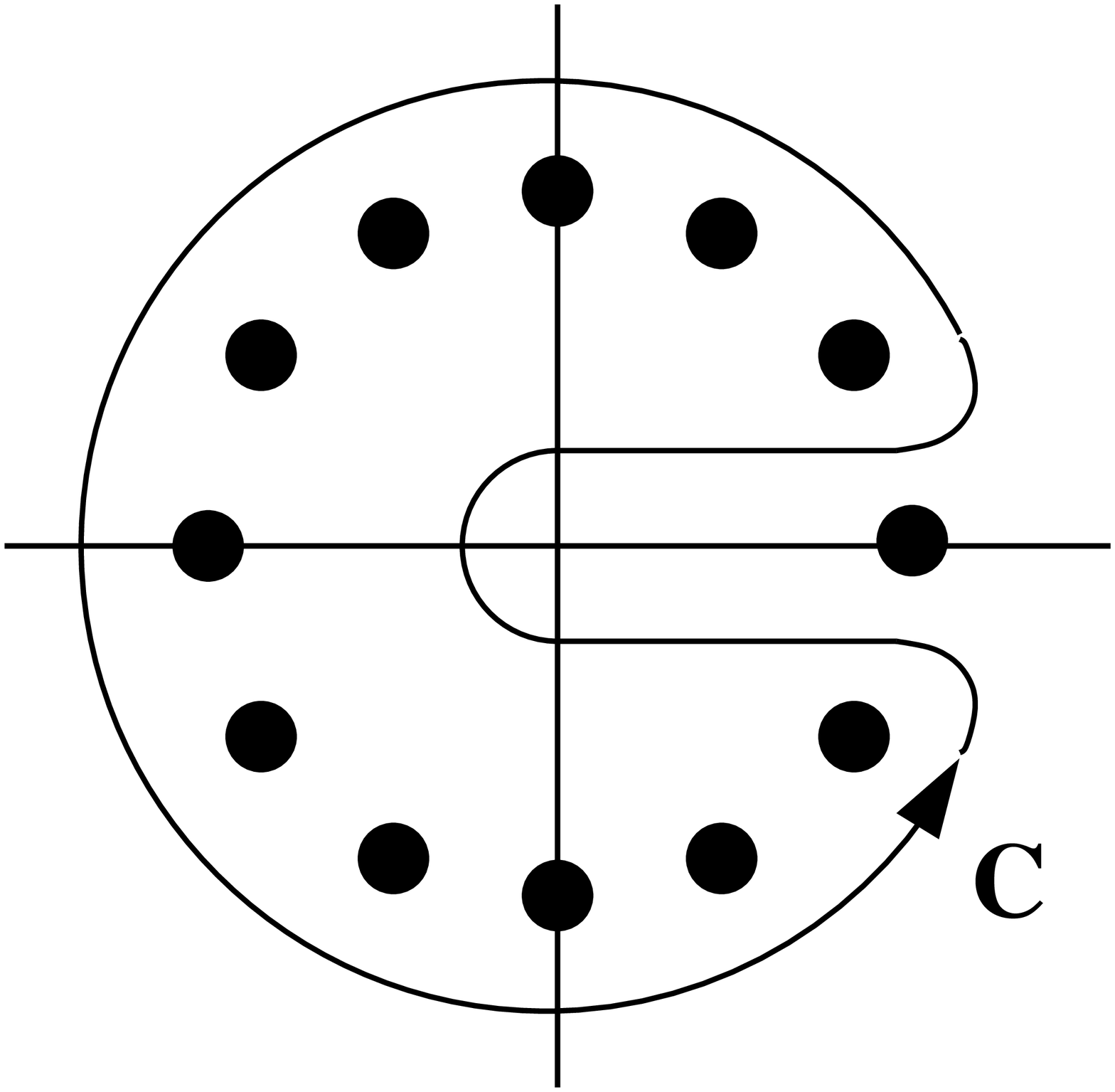}\hspace{2cm}
\includegraphics[scale =0.16]{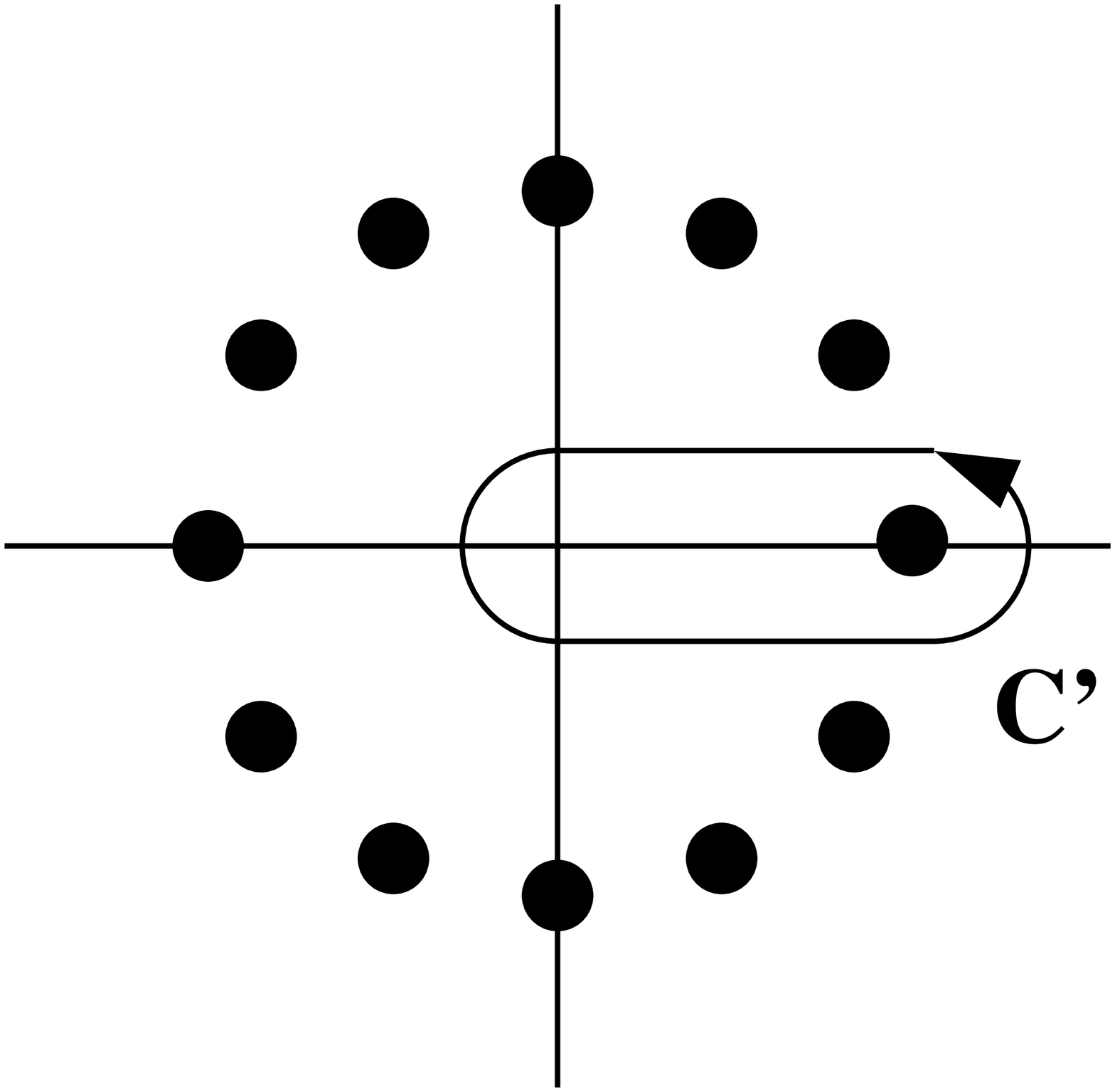}
\end{center}
\caption{Contours for integrations}\label{fig:app-hscontours}
\end{figure}
{\em Proof:} Using Cauchy's theorem~\cite{Lang85} for the function
\begin{displaymath}
  f(z)=\frac{z^m}{z^N(z-1)},\quad 2 \le N,
\end{displaymath}
with the contours drawn in Fig.~\ref{fig:app-hscontours} yields
\begin{eqnarray*}
  \sum_{\alpha=1}^{N-1}\frac{\eta_\alpha^m}{\eta_\alpha -1}
  &=&\frac{1}{2\pi i}\sum_{\alpha=1}^{N-1}
  \oint_C \frac{z^m}{z^N(z-1)}\frac{1}{z-\eta_\alpha} dz \\
  &=&\frac{N}{2\pi i}
  \oint_C \frac{z^{m-1}}{(z-1)(z^N-1)} dz \\
  &=&-\frac{N}{2\pi i}
  \oint_{C'}\frac{z^{m-1}}{(z-1)(z^N-1)} dz \\
  &=&\frac{N+1}{2}-m,
\end{eqnarray*}
where we have used the theorem of residues~\cite{Lang85} in the last step.

\item
\begin{equation}
  \sum_{\alpha=1}^{N-1}
  \frac{\eta_\alpha^m}{\vert\eta_\alpha -1\vert^2}=
  \frac{N^2-1}{12}-\frac{m(N-1)}{2}+\frac{m(m-1)}{2},
  \quad 0\le m \le N.
\label{eq:app-hsfouriersum2}
\end{equation}
\emph{Proof:} We have
\begin{eqnarray*}
\sum_{\alpha=1}^{N-1}\frac{\eta_\alpha^m}{\vert\eta_\alpha -1\vert^2}
&=&-\sum_{\alpha=1}^{N-1}\frac{\eta_\alpha^{m+1}}{(\eta_\alpha -1)^2}\\
&=&-\frac{N}{2\pi i}\oint_C \frac{z^m}{(z-1)^2(z^N-1)} dz \\
&=&\frac{N}{2\pi i}
\oint_{C'}\frac{z^m}{(z-1)^2(z^N-1)} dz \\
&=&\frac{N^2-1}{12}-\frac{m(N-1)}{2}+\frac{m(m-1)}{2}.
\end{eqnarray*}

\item 
For 
\begin{equation}
  A_m=-\sum_{\alpha=1}^{N-1} \eta_\alpha^2 (\eta_\alpha -1)^{m-2}    
\end{equation}
we have:\ \ $A_0=(N-1)(N-5)/12$\ \ by
(\ref{eq:app-hsfouriersum2}),\ \ $A_1=-(N-3)/2$\ \ by
(\ref{eq:app-hsfouriersum1}), and\ \ $A_2=1$\ \ by
$\sum_\alpha^N\eta_\alpha^m=N\delta_{m0}$. Furthermore, 
\begin{eqnarray*}
  A_m&=&-\sum_{\alpha=1}^{N-1} \eta_\alpha^2\sum_{k=0}^{m-2}
  {m-2 \choose k}(-1)^{m-k-2}\eta_\alpha^k\\
  &=&\sum_{k=0}^{m-2}{m-2 \choose k}(-1)^{m-k}
  \Bigl(1-\sum_{\alpha=1}^N \eta_\alpha^{k+2}\Bigr)\\
  &=&\sum_{k=0}^n {n \choose k}(-1)^{n-k}=0,\quad 2<m\le N-1.
\end{eqnarray*}
as the sums of the binomial coefficients of even sites and odd
sites equal each other.

\item For 
\begin{equation}
  B_m=\sum_{\alpha=1}^{N-1} 
  \eta_\alpha (\eta_\alpha+1)(\eta_\alpha-1)^{m-1}
\end{equation}
  we have: $B_0=N-2$, $B_1=-2$, $B_m=0$ for $2<m\le N-2$,
  and $B_{N-1}=N$.
\end{enumerate}

\section{Proof of Theorem~\ref{theo:gstheorem}}\label{app:gstheorem}

As (\ref{eq:theo1}) is symmetric in the $z_i$'s, we prove the theorem
by showing that the LHS of (\ref{eq:theo1}) is independent of $z_1$
and vanishes identically.  To this end, consider the auxiliary
function
\begin{equation}
  \Phi(z_1)=\frac{z_1(z_1-z)^{M-3}}{\prod_{j=2}^M(z_j-z_1)}
  +\sum_{i=2}^M\frac{z_i(z_i-z)^{M-3}}
  {(z_1-z_i)\prod_{j\neq i}^M(z_j-z_i)}.
  \label{eq:appaux6}
\end{equation}
for fixed and distinct $z_2,\ldots,z_M\in\mathbb{C}$.  

First, we show that (\ref{eq:appaux6}) is entire (\ie analytic on
$\mathbb{C}$).  We can restrict ourselves to showing that $\Phi(z_1)$
is analytic at $z_1=z_2$.  (\ref{eq:appaux6}) can be rewritten as
\begin{displaymath}
  \Phi(z_1)=-\frac{z_1(z_1-z)^{M-3}\prod_{j=3}^M(z_j-z_2)
    -z_2(z_2-z)^{M-3}\prod_{j=3}^M(z_j-z_1)}
  {(z_1-z_2)\prod_{j=3}^M(z_j-z_1)(z_j-z_2)}+\tilde{\Phi}(z_1)
\end{displaymath}
where $\tilde{\Phi}$ is analytic at $z_1=z_2$.  Expanding the first
term yields expressions like
\begin{displaymath}
  \frac{1}{z_1-z_2}(z_1^nz_2^m-z_1^mz_2^n),\quad m,n\in\mathbb{N},
\end{displaymath}
which can be shown to be analytic at $z_1=z_2$ by using
(\ref{eq:xysum}).

Second, as the degree of the denominator in (\ref{eq:appaux6}) is
strictly greater than the degree of the numerator, $\Phi(z_1)$
vanishes for $|z_1|\rightarrow\infty$.  Therefore, by Liouville's
theorem~\cite{Lang85} $\Phi$ vanishes for all $z_1\in\mathbb{C}$,
which proves the theorem.

\section{Proof of Theorem~\ref{theo:appsu3-onespinontheorem}}
\label{app:onecolorontheorem}

We prove the theorem in three steps. First, we show that the LHS of
(\ref{eq:appsu3-spinontheorem}) does not depend on the $w_k$'s.
Second, we eliminate the $w_k$'s.  Third, we verify
the remaining equation by explicit calculation.

For the first step, note that (\ref{eq:appsu3-spinontheorem}) is
symmetric under permutations of the $w_k$'s.  It is hence sufficient
to show that (\ref{eq:appsu3-spinontheorem}) does not depend on $w_1$.
Consider for fixed and distinct
$z_1,\dots,z_{M_1},w_2,\dots,w_{M_2}\in\mathbb{C}$ the auxiliary
function
\begin{equation}
\begin{split}
\Phi(w_1)=&\,\frac{1}{z_1-w_1}
\left[w_1\prod^{M_1}_{j=2}\frac{z_j-w_1}{z_j-z_1}
\prod^{M_2}_{l=2}\frac{w_l-z_1}{w_l-w_1}-z_1\right]\\[3mm]
&+\sum_{k=2}^{M_2}\frac{1}{z_1-w_k}
\left[w_k\prod^{M_1}_{j=2}\frac{z_j-w_k}{z_j-z_1}
\prod^{M_2}_{l\neq k}\frac{w_l-z_1}{w_l-w_k}
\cdot\frac{w_1-z_1}{w_1-w_k}-z_1\right]\!.
\end{split}
\label{eq:appsu3-auxiliaryfunction}
\end{equation}
We will show that $\Phi$ is an entire function and bounded for
$|w_1|\rightarrow\infty$.  Then Liouville's theorem~\cite{Lang85}
implies that $\Phi$ is a constant, \ie independent of $w_1$.

To see that $\Phi$ is entire note that
(\ref{eq:appsu3-auxiliaryfunction}) is obviously analytic for
$w_1\in\mathbb{C}\backslash\{z_1,w_2,\dots,w_{M_2}\}$. To
investigate the point $w_1=z_1$, we rewrite the first line of
(\ref{eq:appsu3-auxiliaryfunction}) as
\begin{displaymath}
\frac{1}{z_1-w_1}
\left[w_1\prod^{M_1}_{j=2}\left(1+\frac{z_1-w_1}{z_j-z_1}\right)
\prod^{M_2}_{l=2}\left(1-\frac{z_1-w_1}{w_l-w_1}\right)-z_1\right],
\end{displaymath}
which shows that $\Phi$ can be analytically continued to $w_1=z_1$.
Furthermore, as (\ref{eq:appsu3-auxiliaryfunction}) is symmetric in
$w_2,\dots,w_{M_2}$, we may restrict ourselves now to showing that
$\Phi$ is analytic at $w_1=w_2$. For $w_2=0$ this is obviously the
case. For $w_2\neq 0$ we write $\Phi$ as
\begin{displaymath}
\begin{split}
\Phi (w_1)=&\,\frac{w_1}{z_1-w_1}\prod^{M_1}_{j=2}
\frac{z_j-w_1}{z_j-z_1}\prod^{M_2}_{l=3}\frac{w_l-z_1}{w_l-w_1}
\cdot\frac{w_2-z_1}{w_2-w_1}\\
&+\frac{w_2}{z_1-w_2}\prod^{M_1}_{j=2}\frac{z_j-w_2}{z_j-z_1}
\prod^{M_2}_{l=3}\frac{w_l-z_1}{w_l-w_2}\cdot\frac{w_1-z_1}{w_1-w_2}
+\tilde{\Phi}(w_1),
\end{split}
\end{displaymath}
where $\tilde{\Phi}$ is analytic at $w_1=w_2$. Now,
$\Phi-\tilde{\Phi}$ can be rewritten as
\begin{equation}
\begin{split}
&\frac{1}{(z_1-w_1)(z_1-w_2)}\prod^{M_1}_{j=2}\frac{1}{z_j-z_1}
\prod^{M_2}_{l=3}\frac{w_l-z_1}{(w_l-w_1)(w_l-w_2)}\frac{1}{w_1-w_2}\\
&\qquad\cdot\left[w_1(z_1-w_2)^2\prod^{M_1}_{j=2}(z_j-w_1)
\prod^{M_2}_{l=3}(w_l-w_2)
-w_2(z_1-w_1)^2\prod^{M_1}_{j=2}(z_j-w_2)
\prod^{M_2}_{l=3}(w_l-w_1)\right]\!\!.
\end{split}
\label{eq:appsu3-theoremauxiliary1}
\end{equation}
Expanding the square brackets in
(\ref{eq:appsu3-theoremauxiliary1}) leads to terms like
\begin{displaymath}
\frac{1}{w_1-w_2}(w_1^nw_2^m-w_1^mw_2^n),\quad m,n\in\mathbb{N},
\end{displaymath}
which can be shown to be analytic at $w_1=w_2$ by using
(\ref{eq:xysum}).  This concludes the proof that $\Phi$ is an entire
function of $w_1$.  Furthermore, $\Phi$ tends to a constant for
$|w_1|\rightarrow\infty$ as in the first term of
(\ref{eq:appsu3-auxiliaryfunction}), the degree of the denominator is
greater or equal than the degree of the numerator, while in the second
term the $w_1$-dependence cancels out for $|w_1|\rightarrow\infty$.
Therefore, $\Phi$ is constant by Liouville's theorem~\cite{Lang85}.
As we can replace $z_1$ by any $z_i$ in
(\ref{eq:appsu3-auxiliaryfunction}), the LHS of
(\ref{eq:appsu3-spinontheorem}) does not depend on the $w_k$'s.

In the second step, we eliminate the $w_k$'s from
(\ref{eq:appsu3-spinontheorem}). In order to do so we choose $w_j=z_j$
for $1\le j\le M_1$, and $w_k=Re^{i\varphi_k}$ with distinct
values $\varphi_k\in[0,2\pi]$ for $M_1+1\le k\le M_2$.  We now
have to study three different terms in
(\ref{eq:appsu3-spinontheorem}). First, the terms with $i=k$ yield as
$R\rightarrow\infty$
\begin{displaymath}
-\sum_{i=1}^{M_1}z_i\sum^N_{\gamma=1}(\bar{\eta}_\gamma)^n
\prod^{M_1}_{j\neq i}(\eta_\gamma-z_j).
\end{displaymath}
Second, the terms with $i\neq k, k\le M_1$ yield in the same limit
\begin{displaymath}
-\sum_{i=1}^{M_1}\sum_{k\neq i}^{M_1}\frac{z_i^2}{z_i-z_k}
\sum^N_{\gamma=1}(\bar{\eta}_\gamma)^n
\prod^{M_1}_{j\neq i}(\eta_\gamma-z_j).
\end{displaymath}
Third, the terms with $M_1<k$ vanish in this limit.
Hence (\ref{eq:appsu3-spinontheorem}) reduces
to
\begin{equation}
\begin{split}
\sum_{i=1}^{M_1}z_i\sum^N_{\gamma=1}(\bar{\eta}_\gamma)^n
\prod^{M_1}_{k\neq i}&(\eta_\gamma-z_k)+
\sum_{i=1}^{M_1}\sum_{j\neq i}^{M_1}\frac{z_i^2}{z_i-z_j}
\sum^N_{\gamma=1}(\bar{\eta}_\gamma)^n
\prod^{M_1}_{k\neq i}(\eta_\gamma-z_k)\\
&=-f(n)\sum^N_{\gamma=1}(\bar{\eta}_\gamma)^n
\prod^{M_1}_{i=1}(\eta_\gamma-z_i),
\end{split}
\label{eq:appsu3-theoremauxiliary2}
\end{equation}
with $f(n)=-n(n+1)/2+M_1(M_1+1)/2$. The second term in the LHS of
(\ref{eq:appsu3-theoremauxiliary2}) can be written as
\begin{equation}
\begin{split}
  \sum^N_{\gamma=1}(\bar{\eta}_\gamma)^n&\sum_{i\neq j}^{M_1}
  \frac{z_i^2}{z_i-z_j}(\eta_\gamma-z_j)
  \prod^{M_1}_{k\neq i,j}(\eta_\gamma-z_k)\\
  &=\sum^N_{\gamma=1}
  \Biggl[(\bar{\eta}_\gamma)^{n-1}\sum_{i\neq j}^{M_1}z_i
  -\frac{1}{2}(\bar{\eta}_\gamma)^n\sum_{i\neq j}^{M_1}z_iz_j\Biggr]
  \prod^{M_1}_{k\neq i,j}(\eta_\gamma-z_k),
\end{split}
\label{eq:appsu3-theoremauxiliary3}
\end{equation}
where we have used
\begin{displaymath}
\sum_{i\neq j}^{M_1}\frac{z_i^2}{z_i-z_j}(\eta_\gamma-z_j)=
\frac{1}{2}\sum_{i\neq j}^{M_1}
\bigl[\eta_\gamma(z_i+z_j)-z_iz_j\bigr].
\end{displaymath}

In the third step, we complete the proof by explicit calculation. For
this we use
\begin{equation}
  \sum^N_{\gamma=1}(\bar{\eta}_\gamma)^n
  \prod^{M_1}_{i=1}(\eta_\gamma-z_i)=
  (-1)^{M_1-n}\,\mathcal{S}^{M_1}(z_1\cdots z_{M_1-n}),
\label{eq:calonot}
\end{equation}
where $\mathcal{S}^{M_1}(z_1\cdots z_{M_1-n})$ is the sum over all
possible ways to choose $M_1-n$ coordinates out of
$z_1,\ldots,z_{M_1}$.  Then, using (\ref{eq:appsu3-theoremauxiliary3})
we write (\ref{eq:appsu3-theoremauxiliary2}) as
\begin{equation}
\begin{split}
  &\sum_{i=1}^{M_1}z_i\,\mathcal{S}^{M_1-1}(z_1\cdots z_{M_1-n-1})
  \rule[-8pt]{0.3pt}{18pt}_{\;\mathrm{no}\;z_i}
  +\sum_{i\neq j}^{M_1}z_i\,
  \mathcal{S}^{M_1-2}(z_1\cdots z_{M_1-n-1})
  \rule[-8pt]{0.3pt}{18pt}_{\;\mathrm{no}\;z_i,z_j}\\
  &+\frac{1}{2}\sum_{i\neq j}^{M_1}z_iz_j\,
  \mathcal{S}^{M_1-2}(z_1\cdots z_{M_1-n-2})
  \rule[-8pt]{0.3pt}{18pt}_{\;\mathrm{no}\;z_i,z_j}=f(n)\,
  \mathcal{S}^{M_1}(z_1\cdots z_{M_1-n}).
\end{split}
\label{eq:appsu3-theoremauxiliary4}
\end{equation}
With the relations 
\begin{eqnarray*}
\sum_{i=1}^{M_1}z_i\,
\mathcal{S}^{M_1-1}(z_1\cdots z_{M_1-n-1})
\rule[-8pt]{0.3pt}{18pt}_{\;\mathrm{no}\;z_i}&=&
(M_1-n)\,\mathcal{S}^{M_1}(z_1\cdots z_{M_1-n}),\\
\sum_{i\neq j}^{M_1}z_i\,
\mathcal{S}^{M_1-2}(z_1\cdots z_{M_1-n-1})
\rule[-8pt]{0.3pt}{18pt}_{\;\mathrm{no}\;z_i,z_j}&=&
(M_1-n)n\,\mathcal{S}^{M_1}(z_1\cdots z_{M_1-n}),\\
\sum_{i\neq j}^{M_1}z_iz_j\,
\mathcal{S}^{M_1-2}(z_1\cdots z_{M_1-n-2})
\rule[-8pt]{0.3pt}{18pt}_{\;\mathrm{no}\;z_i,z_j}&=&
(M_1-n)(M_1-n-1)\,\mathcal{S}^{M_1}(z_1\cdots z_{M_1-n}),
\end{eqnarray*}
we can verify (\ref{eq:appsu3-theoremauxiliary4}) and deduce
$f(n)=-n(n+1)/2+M_1(M_1+1)/2$. This concludes the proof of the
theorem.

\section{Proof of Theorem~\ref{theo:twocolorons}}\label{app:twocolorontheorem}

Consider for fixed and distinct
$z_1,\dots,z_{M_1},w_2,\dots,w_{M_2}\in\mathbb{C}$ the auxiliary
function
\begin{equation}
  \begin{split}
    \Phi(w_1)=&\,w_1\prod^{M_1}_{j=2}\frac{z_j-w_1}{z_j-z_1}
    \prod^{M_2}_{l=2}\frac{w_l-z_1}{w_l-w_1}\\[3mm]
    &+\sum_{k=2}^{M_2}w_k\prod^{M_1}_{j=2}\frac{z_j-w_k}{z_j-z_1}
    \prod^{M_2}_{l\neq k}\frac{w_l-z_1}{w_l-w_k}
    \cdot\frac{w_1-z_1}{w_1-w_k}.
  \end{split}
  \label{eq:tc-auxiliaryfunction}
\end{equation}
In analogy to the proof in App.~\ref{app:onecolorontheorem} it can be
shown by using (\ref{eq:xysum}) that (\ref{eq:tc-auxiliaryfunction})
is entire, and that $\Phi$ tends to a constant for
$|w_1|\rightarrow\infty$. By Liouville's theorem~\cite{Lang85} $\Phi$
is constant, and hence (\ref{eq:twocoltheorem}) is independent of the
$w_k$'s.  We can choose the $w_k$'s as $w_j=z_j$ for $1\le j\le M_1$,
and $w_k=Re^{i\varphi_k}$ with distinct values $\varphi_k\in[0,2\pi]$
for $M_1+1\le k\le M_2$. Then, the LHS of (\ref{eq:twocoltheorem})
simplifies to
\begin{equation}
  \sum_{i=1}^{M_1}z_i^2
  \sum^N_{\gamma\delta}(\bar{\eta}_\gamma)^m(\bar{\eta}_\delta)^n
  \prod^{M_1}_{j\neq i}(\eta_\gamma-z_j)(\eta_\delta-z_j).
  \label{eq:tcaux7}
\end{equation}
Using (\ref{eq:calonot}) we can rewrite (\ref{eq:twocoltheorem}) as
\begin{equation}
\begin{split}
&\sum_{i=1}^{M_1}z_i^2\,
\mathcal{S}^{M_1-1}(z_1\cdots z_{M_1-m-1})
\rule[-8pt]{0.3pt}{18pt}_{\;\mathrm{no}\;z_i}\,
\mathcal{S}^{M_1-1}(z_1\cdots z_{M_1-n-1})
\rule[-8pt]{0.3pt}{18pt}_{\;\mathrm{no}\;z_i}\\
&\quad=\,(M_1-m)\,\mathcal{S}^{M_1}(z_1\cdots z_{M_1-m})
\,\mathcal{S}^{M_1}(z_1\cdots z_{M_1-n})\\
&\qquad\;-\sum_{\ell=1}^{\ell_\mathrm{m}}(m-n+2\ell)\,
\mathcal{S}^{M_1}(z_1\cdots z_{M_1-m-\ell})\,
\mathcal{S}^{M_1}(z_1\cdots z_{M_1-n+\ell}).
\end{split}
\label{eq:tcaux8}
\end{equation}

As all terms in (\ref{eq:tcaux8}) are completely symmetric in the
$z_i$'s, we simplify the notation by imagining to have written an
operator $\mathcal{S}^{M_1}$ around each term and then order all terms
inside, \ie we replace
\begin{displaymath}
  \mathcal{S}^{M}(z_1\cdots z_{m})\rightarrow 
  {M\choose m}\,z_1\cdots z_m.
\end{displaymath}
The remainder of the proof makes use of the auxiliary theorem
\begin{lemma}
\begin{equation}
  \label{eq:lemma}
  \mathcal{S}^{M}(z_1\cdots z_{m})\mathcal{S}^{M}(z_1\cdots z_{n})
  \rightarrow\sum_{k=k_\mathrm{min}}^{k_\mathrm{max}} {M\choose m}
  {m\choose k}{M-m\choose n-k}\,[k,m+n-k],
\end{equation}
where $k_\mathrm{min}=\mathrm{max}(0,m+n-M)$,
$k_\mathrm{max}=\mathrm{min}(m,n)$, and $[k,m+n-k]=z_1^2\cdots
z_k^2\,z_{k+1}\cdots z_{m+n-k}$.
\end{lemma}
\emph{Proof:} Replace first the LHS as
\begin{displaymath}
\mathcal{S}^{M}(z_1\cdots z_{m})\mathcal{S}^{M}(z_1\cdots
z_{n})\rightarrow{M\choose m}\,z_1\cdots z_m\;\mathcal{S}^{M}(z_1\cdots
z_{n}).
\end{displaymath}
In order to obtain the contribution $z_1^2\cdots z_k^2$, $k$
coordinates out of $z_1,\ldots,z_n$ have to match $k$ coordinates out
of $z_1,\ldots,z_m$, which yields ${m\choose k}$ terms for
$k_\mathrm{min}\le k\le k_\mathrm{max}$. The remaining $n-k$
coordinates out of $z_1,\ldots,z_n$ match $z_{m+1},\ldots,z_{m+n-k}$,
which yields ${M-m \choose n-k}$ terms.\hfill$\Box$

\vspace{3mm}
Now, by application of this lemma the first term on the RHS of
(\ref{eq:tcaux8}) simplifies to
\begin{displaymath}
  (M_1-m)\sum_{k=k_\mathrm{m}}^{M_1-m} {M_1\choose m}
  {M_1-m\choose k}{m\choose M_1-n-k}\,[k,2M_1-m-n-k],
\end{displaymath}
where $k_\mathrm{m}=\mathrm{max}(0,M_1-m-n)$ and we have used $n\le
m$.  The second term on the RHS of (\ref{eq:tcaux8}) reads
\begin{displaymath}
-\sum_{\ell=1}^{\ell_\mathrm{m}}
\sum_{k=k_\mathrm{m}}^{M_1-m-\ell}\!\!(m-n+2\ell)\,
{M_1\choose m+\ell}
{M_1-m-\ell\choose k}{m+\ell\choose M_1-n+\ell-k}\,[k,2M_1-m-n-k].
\end{displaymath}
In this equation we change the order of the summations according to
\begin{displaymath}
\sum_{\ell=1}^{\ell_\mathrm{m}}\,\sum_{k=k_\mathrm{m}}^{M_1-m-\ell}
\rightarrow
\sum_{k=k_\mathrm{m}}^{M_1-m-1}\,\sum_{\ell=1}^{M_1-m-k}.
\end{displaymath}
On the LHS of (\ref{eq:tcaux8}) at least $z_1^2$
is present and hence we find 
\begin{displaymath}
  M_1\sum_{k=k'_\mathrm{m}}^{M_1-m} {M_1-1\choose m}
  {M_1-m-1\choose k-1}{m\choose M_1-n-k}\,[k,2M_1-m-n-k],
\end{displaymath}
where $k'_\mathrm{m}=\mathrm{max}(1,M_1-m-n)$. By considering each $k$
separately, (\ref{eq:tcaux8}) simplifies to
\begin{equation}
\begin{split}
  \frac{k}{(M_1-m-k)!(M_1-n-k)!}&=
  \frac{M_1-m}{(M_1-m-k)!(M_1-n-k)!}\\
  &-\sum_{\ell=1}^{M_1-m-k}
  \frac{m-n+2\ell}{(M_1-m-k-\ell)!(M_1-n-k+\ell)!}.
\end{split}
\label{eq:last}
\end{equation}
For $k=M_1-m$, (\ref{eq:last}) is trivially satisfied.  For
$\mathrm{max}(0,M_1-m-n)\le k\le M_1-m-1$, (\ref{eq:last})
reduces to
\begin{displaymath}
\frac{m}{m!\,n!}=\sum_{\ell=1}^{m}\frac{n-m+2\ell}{(m-\ell)!(n+\ell)!},
\end{displaymath}
which is easily verified.  This completes the proof of
Theorem~\ref{theo:twocolorons}.


\begin{thebibliography}{10}

\bibitem{IsobeUeda96}
M. Isobe and Y. Ueda, J. Phys. Soc. Jpn. {\bf 65},  1178  (1996).

\bibitem{FujiiNakaoYosihamaNishiNakajimaKakuraiIsobeUedaSawa97}
Y. Fujii, H. Nakao, T. Yosihama, M. Nishi, K. Nakajima, K. Kakurai, M. Isobe,
  Y. Ueda, and H. Sawa, J. Phys. Soc. Jpn. {\bf 66},  326  (1997).

\bibitem{KitaokaKobayashiKodaWakabayashiNiinoYamakageTaguchiAmayaYamauraTakano%
HiranoKanno98}
Y. Kitaoka, T. Kobayashi, A. K\={o}da, H. Wakabayashi, Y. Niino,
H. Yamakage, S. Taguchi, K. Amaya, K. Yamaura, M. Takano,
A. Hirano, and R. Kanno, J. Phys. Soc. Jpn. {\bf 67},  3703  (1998).

\bibitem{TokuraNagaosa00}
Y. Tokura and N. Nagaosa, Science {\bf 288},  462  (2000).

\bibitem{LiMaShiZhang98}
Y.~Q. Li, M. Ma, D.~N. Shi, and F.~C. Zhang, Phys. Rev. Lett. {\bf 81},  3527
  (1998).

\bibitem{FrischmuthMilaTroyer99}
B. Frischmuth, F. Mila, and M. Troyer, Phys. Rev. Lett. {\bf 82},  835  (1999).

\bibitem{MilaFrischmuthDeppelerTroyer99}
F. Mila, B. Frischmuth, A. Deppeler, and M. Troyer, Phys. Rev. Lett. {\bf 82},
  3697  (1999).

\bibitem{AzariaGogolinLecheminantNerseyan99}
P. Azaria, A.~O. Gogolin, P. Lecheminant, and A.~A. Nersesyan, Phys. Rev. Lett.
  {\bf 83},  624  (1999).

\bibitem{BosscheAzariaLecheminantMila01}
M. van den Bossche, P. Azaria, P. Lecheminant, and F. Mila, Phys. Rev. Lett.
  {\bf 86},  4124  (2001).

\bibitem{HonerkampHofstetter04}
C. Honerkamp and W. Hofstetter, Phys. Rev. Lett. {\bf 92},  170403  (2004).

\bibitem{AssarafAzariaBoulatCaffarelLecheminant04}
R. Assaraf, P. Azaria, E. Boulat, M. Caffarel, and P. Lecheminant, Phys. Rev.
  Lett. {\bf 93},  016407  (2004).

\bibitem{AbrahamMcAlexanderGertonHuletCoteDalgarno97}
E.~R.~I. Abraham, W.~I. McAlexander, J.~M. Gerton, R.~G. Hulet, R. {C\^ot\'e},
  and A. Dalgarno, Phys. Rev. A {\bf 55},  R3299  (1997).

\bibitem{RegalJin03}
C.~A. Regal and D.~S. Jin, Phys. Rev. Lett. {\bf 90},  230404  (2003).

\bibitem{SchurichtGreiter05colepl}
D. Schuricht and M. Greiter, Europhys. Lett. {\bf 71}, 987 (2005).

\bibitem{Haldane88}
F.~D.~M. Haldane, Phys. Rev. Lett. {\bf 60},  635  (1988).

\bibitem{Shastry88}
B.~S. Shastry, Phys. Rev. Lett. {\bf 60},  639  (1988).

\bibitem{Haldane91prl1}
F.~D.~M. Haldane, Phys. Rev. Lett. {\bf 66},  1529  (1991).

\bibitem{HaldaneHaTalstraBernardPasquier92}
F.~D.~M. Haldane, Z.~N.~C. Ha, J.~C. Talstra, D. Bernard, and V. Pasquier,
  Phys. Rev. Lett. {\bf 69},  2021  (1992).

\bibitem{Kawakami92prb1}
N. Kawakami, Phys. Rev. B {\bf 46},  1005  (1992).

\bibitem{Kawakami92prb2}
N. Kawakami, Phys. Rev. B {\bf 46},  R3191  (1992).

\bibitem{HaHaldane92}
Z.~N.~C. Ha and F.~D.~M. Haldane, Phys. Rev. B {\bf 46},  9359  (1992).

\bibitem{HaHaldane93} 
Z.~N.~C. Ha and F.~D.~M. Haldane, Phys. Rev. B {\bf 47}, 12459 (1993).

\bibitem{Wilczek1990}
F. Wilczek, {\em Fractional statistics and anyon superconductivity} 
(World Scientific, 1990).

\bibitem{Haldane91prl2}
F.~D.~M. Haldane, Phys. Rev. Lett. {\bf 67},  937  (1991).

\bibitem{Georgi82}
H. Georgi, {\em Lie Algebras in Particle Physics} 
(Addison-Wesley, Redwood City, 1982).

\bibitem{ChariPressley98}
V. Chari and A. Pressley, {\em A {G}uide to {Q}uantum {G}roups} 
(Cambridge University Press, Cambridge, 1998).

\bibitem{BouwknegtSchoutens96}
P. Bouwknegt and K. Schoutens, Nucl. Phys. {\bf B482},  345  (1996).

\bibitem{Schoutens97}
K. Schoutens, Phys. Rev. Lett. {\bf 79},  2608  (1997).

\bibitem{YamamotoSaigaArikawaKuramoto00prl} 
T.~Yamamoto, Y.~Saiga, M.~ Arikawa, and Y.~Kuramoto,
Phys. Rev. Lett. {\bf 84}, 1308 (2000).

\bibitem{YamamotoSaigaArikawaKuramoto00jpsj} 
T.~Yamamoto, Y.~Saiga, M.~ Arikawa, and Y.~Kuramoto, 
J. Phys. Soc. Jpn. {\bf 69}, 900 (2000).

\bibitem{KuramotoKato95} 
Y. Kuramoto and Y. Kato, J. Phys. Soc. Jpn. {\bf 64}, 4518 (1995).
  
\bibitem{KatoKuramoto96} 
Y. Kato and Y. Kuramoto, J. Phys. Soc. Jpn. {\bf 65}, 1622 (1996).
  
\bibitem{BouwknegtSchoutens99}
P. Bouwknegt and K. Schoutens, Nucl. Phys. {\bf B547},  501  (1999).

\bibitem{manuscriptinpreparationGS}
M. Greiter and D. Schuricht, manuscript in preparation.

\bibitem{BernevigGiulianoLaughlin01prb}
B.~A. Bernevig, D. Giuliano, and R.~B. Laughlin, Phys. Rev. B {\bf 64},  24425
  (2001).
  
\bibitem{Sutherland71pra} B. Sutherland, Phys. Rev. A {\bf 4}, 2019
  (1971).
  
\bibitem{Sutherland72} B. Sutherland, Phys. Rev. A {\bf 5}, 1372
  (1972).
  
\bibitem{GreiterSchurichtsiprb}
M. Greiter and D. Schuricht, Phys. Rev.~B {\bf 71}, 224424 (2005).

\bibitem{Essler95}
F.~H.~L. E{\ss}ler, Phys. Rev. B {\bf 51},  13357  (1995).

\bibitem{manuscriptinpreparationGST}
M. Greiter, D. Schuricht, and R. Thomale, manuscript in preparation.

\bibitem{Andrei92}
N. Andrei,  in {\em Low-dimensional quantum field theories for condensed matter
  physicists}, edited by S. Lundquist, G. Morandi, and Y. Lu (World Scientific,
  Singapore, 1992).

\bibitem{Lang85}
S. Lang, {\em Complex Analysis} (Springer, New York, 1985).

\end{thebibliography}
\end{document}